%% file: mb_mwj_nw.tex
\documentclass[a4paper,12pt]{article}
\usepackage{fullpage,graphicx,latexsym,mathtools,amsmath,amssymb,dsfont,authblk,natbib,appendix}
\usepackage[font=scriptsize,labelfont=bf]{caption}
\usepackage[labelformat=simple]{subcaption}
\newcommand{\defeq}{\vcentcolon=}
\newcommand{\lb}{\left(}
\newcommand{\rb}{\right)}
\newcommand{\lcb}{\left\{}
\newcommand{\rcb}{\right\}}

\DeclareMathOperator*{\argmin}{arg\,min}
\DeclareMathOperator*{\argmax}{arg\,max}
\begin{document}
\sloppy
\lefthyphenmin=2
\righthyphenmin=2

\title{A hidden Markov model for decoding and the analysis of replay in spike trains}

\author[1]{Marc Box\thanks{mb0184@bristol.ac.uk}}
\author[2]{Matt W. Jones\thanks{matt.jones@bristol.ac.uk}}
\author[3]{Nick Whiteley\thanks{nick.whiteley@bristol.ac.uk}}

\affil[1]{Bristol Centre for Complexity Sciences, University of Bristol}
\affil[2]{School of Physiology and Pharmacology, University of Bristol}
\affil[3]{School of Mathematics, University of Bristol}


\maketitle

\begin{abstract}
We present a hidden Markov model that describes variation in an animal's position associated with varying levels of activity in action potential spike trains of individual place cell neurons. The model incorporates a coarse-graining of position, which we find to be a more parsimonious description of the system than other models. We use a sequential Monte Carlo algorithm for Bayesian inference of model parameters, including the state space dimension, and we explain how to estimate position from spike train observations (decoding). We obtain greater accuracy over other methods in the conditions of high temporal resolution and small neuronal sample size. We also present a novel, model-based approach to the study of replay: the expression of spike train activity related to behaviour during times of motionlessness or sleep, thought to be integral to the consolidation of long-term memories. We demonstrate how we can detect the time, information content and compression rate of replay events in simulated and real hippocampal data recorded from rats in two different environments, and verify the correlation between the times of detected replay events and of sharp wave/ripples in the local field potential.
\end{abstract}

\input{introduction.tex}
\input{methods.tex}
\input{results.tex}
\input{discussion.tex}
\appendixpage
\appendix
\input{appendices.tex}

\section*{Acknowledgements}
Our thanks to Nadine Becker and Josef Sadowski for generous sharing of spike trains. MB is grateful to the Bristol Centre for Complexity Sciences for their support and for funding through EPSRC (EP/I013717/1). MWJ would like to acknowledge the BBSRC (BBG006687) and MRC (G1002064) for financial support.


\input{mb_mwj_nw.bbl}
\end{document}

%% file: introduction.tex
\section{Introduction}
\label{introduction}
\subsection{Background and motivation}
\label{subsec:background}
This article is concerned with the development of statistical modelling techniques for multiple concurrent spike trains recorded from behaving rats using implanted microelectrodes. We are interested in data sets that include other variables, for example position in a maze, that may be correlated with concurrent spike trains. We focus on two applications relevant to this context: the \emph{decoding} of position information encoded in hippocampal spike trains and the detection and analysis of spike train \emph{replay}.

\subsubsection{Decoding}
\label{subsubsec:decodingintroduction}
Decoding is the task of estimating the information content transmitted by spike trains: sequences of times of spikes, or action potentials, recorded from individual neurons and considered as instantaneous and identical events (\cite{Rieke1999}). Decoding has been used for the study of place cells: pyramidal cells of the hippocampus that spike selectively in response to the animal's position (\cite{O'Keefe1971}, \cite{O'Keefe1976}). Individual cells have been observed to encode collectively entire environments in this manner (``population coding'' of space, \cite{Wilson1993}). With large scale, parallel microelectrode recordings (\cite{Buzsaki2004}) it is possible to accurately decode the trajectory of an animal around an environment from population activity, with increasing accuracy as more cells are sampled (\cite{Zhang1998}). In this article, \emph{position} is the variable of interest for encoding and decoding, but these ideas can be applied more generally to other sensory or behavioural variables.

\subsubsection{Replay}
\label{subsubsec:replayintroduction}
\emph{Replay} is the reoccurrence of population spiking activity associated with a specific stimulus (an association made \emph{online}: when the stimulus was presented), during times of unrelated behaviour (\emph{offline}: times of sleep or motionlessness). The phenomenon has been most extensively studied in the place cells of rodents, in which spike trains encoding the trajectory of the animal are replayed in this manner. The time of hippocampal replay events has been found to correlate with the time of local field potential (LFP) events known as sharp wave/ripples (SWR, \cite{Buzsaki1992}), by \cite{Foster2006}, \cite{Diba2007a} and \cite{Davidson2009} during awake restful behaviour, and by \cite{Kudrimoti1999} during sleep.

Place cell replay has been demonstrated to occur on a faster timescale than the encoded trajectory: 20 times faster for cells of the hippocampus (\cite{Nadasdy1999}, \cite{Lee2002a}) and 5 to 10 times faster for cells of the cortex (\cite{Ji2006}, \cite{Euston2007}). In the hippocampus this compression of spiking activity may be due to the burst firing of cells induced by SWR events (\cite{Csicsvari1999}), or the coordination of place cells by the LFP theta rhythm (\cite{O'Keefe1993}), but it is not clear what is responsible for the effect in the cortex (\cite{Buhry2011}).

Although replay, and in particular \emph{preplay} - the expression of offline behavioural sequences \emph{prior} to the behaviour (\cite{Diba2007a}, \cite{Dragoi2011}) - have been suggested to play a role in active cognitive processes (\cite{Gupta2010}, \cite{Pfeiffer2013}), most of the literature concerned with the role of replay has focussed on the \emph{consolidation hypothesis} (\cite{O'Neill2010}, \cite{Carr2011}): that experiences are encoded online by cell assemblies in the hippocampus, then transmitted to the cortex for long-term storage during offline replay. This is supported by observations that hippocampal SWR coincide with high frequency oscillations in the cortex (\cite{Siapas1998}, \cite{Molle2006}), by observations of coordinated activation of cortical cells during hippocampal replay (\cite{Ji2006}, \cite{Euston2007}, \cite{Peyrache2009}), and by slowing of learning by blocking SWRs (\cite{Girardeau2009}, \cite{Ego2010}). Furthermore, correlated offline spiking patterns between pairs of cells within and between the hippocampus and cortex has been observed by \cite{Qin1997} and \cite{Sutherland2000}. However, it remains to be demonstrated whether the same encoded information is being replayed within the two regions during replay events, as implied by the consolidation hypothesis.

\subsection{Current model-based approaches to decoding and replay detection}
\label{subsec:currentapproaches}
A simple statistical model used for decoding was described by \cite{Zhang1998} and compared favourably with nonparametric methods. This model, which we will refer to as the \emph{Bayesian decoder (BD)}, has been influential in spike train analysis in general (\cite{Chen2013}) and replay analysis in particular (e.g. in \cite{Davidson2009}, \cite{Karlsson2009}, \cite{Dragoi2011}, \cite{Pfeiffer2013}, and \cite{Wikenheiser2013}). It consists of a parametric model for the number of spikes in consecutive time intervals, with position encoded as the expected spike count in each interval. Parameter values are estimated from a data set of observed spike trains and position using the method of maximum likelihood, and decoding is achieved by positing a prior distribution for position and using Bayes' theorem to derive the posterior distribution over position given spike train observations. The BD approach to decoding is used as a performance benchmark in Section \ref{subsec:decodingresults}.

Replay has previously been detected as the improved correlation of cell pair firing rates post-behaviour by \cite{Pavlides1989}, \cite{Wilson1994}, and \cite{Skaggs1996}, and by using pattern-matching techniques in spike trains by \cite{Nadasdy1999} and \cite{Louie2001}. More recently, statistical model-based decoding techniques such as BD have allowed researchers to begin to ask questions about replay directly in terms of the observable that is supposed to be encoded rather than purely as a spike train phenomenon: whether replay is preferentially of trajectories of a certain length, complexity or location, for example.

More complex models have attempted to account for the strong dependence through time of processes such as the trajectory of an animal and its concurrent spike trains in order to achieve greater accuracy of representation and decoding. In the state space model of \cite{Brown1998}, and in the hidden Markov model (HMM) of \cite{Johnson2007}, spike counts are conditionally independent observations given the position, which constitutes a latent process. That is, a Markovian dependence structure is assumed for the position process, characterised by a transition matrix and initial state distribution. The spike train model is identical to that of BD. We refer to this model as the \emph{latent position (LP)} hidden Markov model.

In the application of the HMM presented in \cite{Johnson2007}, the state space is determined by the set of positions explored, which may constitute far greater model complexity than is sufficient to characterise the spike train observations, thus incurring a greater computational burden and requiring more data in order to estimate the extra parameters. In \cite{Chen2012}, a HMM is employed in which the state space is not identified with the set of positions (but is interpreted as a ``virtual environment''). Parameters of the Markov chain are estimated from spike train observations only, rather than direct observations of the hidden process as in \cite{Johnson2007}. The number of states required to sufficiently characterise observations is determined through a process of model selection. Thus, \cite{Chen2012} are able to elicit directly from a spike train ensemble the distinct patterns of activity in place cells that may encode position, without needing to prespecify the receptive fields of these cells (the \emph{place fields}, as would be necessary in a nonparametric approach), and to infer from the transition matrix the ``topology'' of the spatial representation.

\subsection{The contributions of this article}
\label{subsec:contributionofarticle}
\paragraph{Model relating place cell spike trains to position}
We present a statistical model, the \emph{observed position (OP)} model, that offers improved performance for decoding and for the study of replay over the BD and LP models. Like \cite{Chen2012} we posit a HMM structure with an unobserved latent process to characterise the variation in observed processes. The difference in our model from \cite{Chen2012} is that we represent position as an observation process in parallel to the spike trains, allowing us to perform decoding when position data is missing, as in BD and LP. We find with the OP model that we achieve better performance in decoding than the BD and LP models when we use a high time resolution and when we have spike trains from a small number of cells.

\paragraph{A Bayesian inference algorithm for parameters and model size}
We make use of a sequential Monte Carlo (SMC) algorithm to perform Bayesian parameter inference, with a state space transformation suggested by \cite{Chopin2007} to make the HMM identifiable. This algorithm makes a numerical approximation (achieving greater accuracy with larger SMC sample size) to the exact posterior distribution over parameters (the variational Bayes method used by \cite{Chen2012} only targets approximate values of parameters). Our algorithm also makes simultaneous inference for the number of states of the model.

\paragraph{New methods for the analysis of replay}
We also introduce a new model-based technique for the reliable detection of the replay of specific trajectories on different time scales. We are able to compare the times of replay events for particular trajectories that may vary in spatial characteristics, duration and compression in time relative to behaviour. These properties of our methods make them useful in particular for exploring evidence that the information content of replay is coordinated between different neuronal populations, such as the hippocampus and neocortex.

\subsection{Structure of the article}
Section \ref{sec:methods} describes our data (Section \ref{subsec:data}) and our model (Sections \ref{subsec:modelling} and \ref{subsec:priorsandfullconditionals}), explains how we perform inference for model parameters, hidden states, and missing position data (decoding) (Section \ref{subsec:inferenceinmodel}), and explains the analysis of replay within our model, including inference for the time and content of replay (Section \ref{subsec:replaydetection}). Also is explained how we detect SWR events and demonstrate correlation with replay events using the cross correlogram (Section \ref{subsec:replayripplecorrelation}) and the simulation of data (Section \ref{subsec:simulation}). Section \ref{sec:results} presents results from applying our model to simulated and real (experiment-generated) data. Model fitting results which demonstrate the model's characterisation of spike train and position data are presented (Section \ref{subsec:fittingresults}), also the results of decoding position comparing our model against the BD and LP alternatives (Section \ref{subsec:decodingresults}), and our analysis of replay in simulated and real sleep data (Section \ref{subsec:replayanalysisresults}). These results are discussed, and our methods appraised, in Section \ref{sec:discussion}.

%% file: methods.tex
\section{Methods}
\label{sec:methods}
\subsection{Description of the experimental data}
\label{subsec:data}
Our experimental data sets consist of simultaneous recordings of a rat's position and hippocampal spike trains. Two environments were used: a straight linear track and a double-ended T-maze (see \cite{Jones2005} for details). In each of these, a rat performed repeated consecutive trials of a reinforced learning task. In the linear track this consists in running from one end to the other, where food reward is received. In the T-maze the rat runs between rest sites in the terminal ends of corridors on opposite sides of the maze. Food reward is received at these sites, but on one side of the maze only when the correct corridor away from the ``T'' junction is chosen, reliably determined by recent experience.

In both experimental setups, two epochs of different behavioural conditions were used: a \emph{RUN} epoch, in which the animal performed the learning task in the environment, immediately followed by a \emph{REST} epoch, in which the animal remained in a separate dark box, in a state of quiescence likely including sleep. Spike trains were recorded from up to $19$ hippocampal place cells throughout both epochs, and position in the environment was recorded using an infrared camera. Thus, for each environment we have a RUN data set (of spike trains and position) which we use for model parameter inference and for decoding analysis, and a REST data set (of spike trains only) which we use for replay analysis.

\subsection{Modelling}
\label{subsec:modelling}
This section describes the OP model: a parametric model for discretised spike trains and position observations related via a hidden discrete time Markov chain. The model structure and parameterisation are explained in Sections \ref{subsubsec:hmm} and \ref{subsubsec:model}. Section \ref{subsubsec:augmentedstates} addresses the identifiability of model parameters.

\subsubsection{Data discretisation}
\label{subsubsec:discretisation}
Our spike train data consists of observations from $C$ distinct point processes in continuous time. We use a time interval width $\delta t$ seconds to partition this data into $T$ time bins, and we let $Y_{t, n}$ for $1 \le n \le C$ and $1 \le t \le T$ represent the number of times neuron $n$ spikes in the $t^\text{th}$ time bin. We denote the random vector of spike counts from each neuron at time $t$ as $\mathbf{Y}_t$, and we denote a time vector of variables between time bins $t_1$ and $t_2$ inclusive as $\mathbf{Y}_{t_1:t_2}$. We use the lowercase, as in $\mathbf{y}_{t_1:t_2}$, to represent observed spike counts.

We use $X_t$, for $1 \le t \le T$, to denote the random discrete position of the animal in time bin $t$. Our position data consists of a sequence of two dimensional pixel coordinates recorded at a frequency of $25$Hz. This will exceed any frequency implied by $\delta t$ we use; therefore we can easily adapt these data to our discrete time scale of $T$ time bins by taking the first observation in each bin.

We discretise space so that each $X_t$ is a finite random variable. The raw two dimensional pixel coordinates are partitioned into a square grid; we then mark as inaccessible all grid squares covering regions outside of the maze. The remaining squares we label arbitrarily from $1$ to $M$, forming the domain of $X_t$.

\subsubsection{HMM to relate spike trains to position}
\label{subsubsec:hmm}
\begin{figure}[h]
\centering
\includegraphics[scale=0.3]{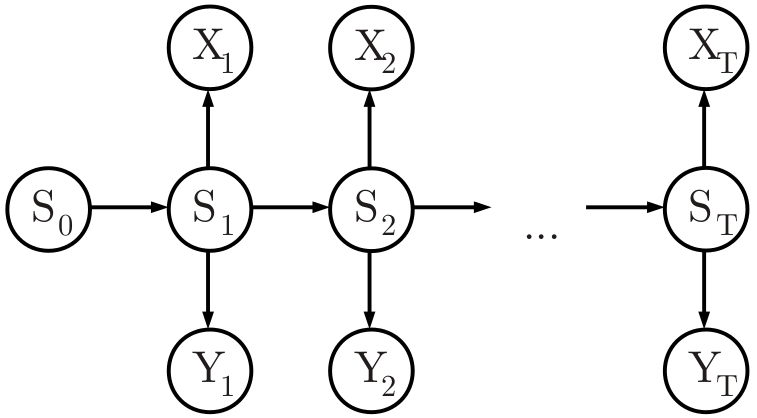}
\caption{DAG for the LP model, explained in Section \ref{subsec:modelling}.}
\label{fig:dag}
\end{figure}
We posit a discrete time Markov chain with $\kappa$ states underlying the observation processes, denoted $S_{0:T}$, with transition matrix $\mathbf{P} = \lb P_{i, j} \rb$ where $P_{i, j} \defeq Pr\lb S_t = j \mid S_{t - 1} = i \rb$ for $1 \le i, j \le \kappa$ and for all $1 \le t \le T$, and initial state distribution $\mathbf{\pi} = \lb \pi_i \rb$ where $\pi_i \defeq Pr\lb S_0 = i \rb$ for $1 \le i \le \kappa$. The dependence between observation variables and the Markov chain is depicted in the directed acyclic graph (DAG) of Fig. \ref{fig:dag}.

We assume $Y_{t, n}$ and $X_t$ are conditionally independent of $\mathbf{Y}_{1:t - 1, n}, \mathbf{Y}_{t + 1:T, n}, X_{1:t - 1}, X_{t + 1:T}, S_{0:t - 1}$ and $S_{t + 1:T}$ for each $t$, given $S_t$, so the joint probability of all model variables factorises as
\begin{equation}
p \lb   \mathbf{y}_{1:T},   x_{1:T},   s_{0:T}   \mid   \mathbf{\theta},   \kappa   \rb
=   \pi_{s_0}   \prod_{t = 1}^T   p \lb   \mathbf{y}_t,   x_t   \mid   s_t,   \mathbf{\theta},   \kappa   \rb   P_{s_{t - 1}, s_t},
\end{equation}
in which $\mathbf{\theta}$ represents the set of all model parameters. We further assume the conditional independence of $Y_{1:T, n}$ for spike trains $1 \le n \le C$ and positions $X_{1:T}$ given $S_{1:T}$, so the likelihood factorises as
\begin{equation}
p \lb   \mathbf{y}_t,   x_t   \mid   s_t,   \mathbf{\theta},   \kappa   \rb
=   p \lb   x_t   \mid   s_t,   \mathbf{\theta},   \kappa   \rb   \prod_{n = 1}^C   p \lb   y_{t, n}   \mid   s_t,   \mathbf{\theta},   \kappa   \rb.
\end{equation}

\subsubsection{Parametric observation models}
\label{subsubsec:model}
\paragraph{Spike trains}
We model our discrete spike trains $Y_{1:T, n}$ as Poisson random variables with piecewise constant means and with jumps between means on changes of state of the Markov chain. That is, we posit $\kappa$ distinct Poisson rates for each spike train, denoted $\lambda_{i, n}$ for $1 \le i \le \kappa$ and $1 \le n \le C$. Thus $Y_{t, n} \mid S_t = s \sim \mathtt{Poi}\lb \delta t \lambda_{s, n} \rb$, and
\begin{equation}
\label{eq:poissonlikelihood}
p \lb   Y_{t, n} = y_{t, n}   \mid   S_t = i,   \mathbf{\theta},   \kappa   \rb   =   e^{-\delta t   \lambda_{i, n}}   \frac{   \lb   \delta t   \lambda_{i, n}    \rb ^ {y_{t, n}}   }{   y_{t, n}!   }.
\end{equation}

\paragraph{Position}
We model $X_t$ using $\kappa$ distinct categorical distributions, labelled by $S_t$, over the set of outcomes $\{1, 2, \dots, M\}$ that jump in parallel with the spike train processes. Outcomes of the $i^\text{th}$ distribution are explained by an underlying two dimensional Gaussian with mean $\xi_i$ and covariance matrix $\Sigma_i$. These are the only free parameters of the position model.

This is achieved by mapping discrete positions $1$ to $M$ to the Euclidean plane using a transformation that preserves the topology of the maze, as follows. We define a distance function $d: \{ 1, 2, \dots, M \} \times \{ 1, 2, \dots, M \} \to \mathbb{R}$ that returns the distance between two positions when access from one to the other is constrained to traversable maze regions (i.e. along corridors). This is achieved by measuring the distance cumulatively through adjacent positions; see Appendix \ref{app:distance} for details. We use the transformation $\mathbf{f}_x: \{1, 2, \dots, M\} \to \mathbb{R}^2$ to map discrete positions $x'$ to vectors in $\mathbb{R}^2$ of length $d(x, x')$ and bearing from the origin equal to the true bearing of $x'$ from $x$ (measured from the centres of the grid squares demarking these positions). The categorical probabilities for our discrete position model are then
\begin{equation}
\label{eq:positionprob}
p \lb   X_t = x   \mid   S_t = i,   \xi_i,   \Sigma_i   \rb
=   \frac{q \lb   \mathbf{f}_{\xi_i} \lb   x   \rb;   0,   \Sigma_i   \rb}{\sum_{x' = 1}^M   q\lb   \mathbf{f}_{\xi_i}   \lb   x'   \rb;   0,   \Sigma_i   \rb},
\end{equation}
where
\begin{equation}
\label{eq:gaussianpdf}
q \lb   \mathbf{f}_{\xi_i}   \lb   x   \rb;   0,   \Sigma_i   \rb
=   \exp \lcb   \mathbf{f}_{\xi_i}   \lb   x   \rb^\intercal   \Sigma^{-1}_i   \mathbf{f}_{\xi_i}   \lb   x   \rb   \rcb,
\end{equation}
the unnormalised probability density of the two dimensional Gaussian distribution with mean $0$ and covariance matrix $\Sigma_i$ evaluated at $\mathbf{f}_{\xi_i}   \lb   x   \rb$.

The purpose of this general approach is that we obtain a position model that satisfies our intuition for the accessibility of places from each other in non-convex environments such as a T-maze. In particular the distribution over $X_t$ gven a particular state should be unimodal, having monotonically decreasing probability with distance from the modal position, since positions of similar probability should be local. This is violated in a concave environment when using the Euclidean distance in place of $d$.

By thus constraining the categorical outcome probabilities, we reduce the number of free parameters from $M - 1$ for each state to simply a modal position $\xi_i$ and a covariance matrix $\Sigma_i$ for each state. Therefore, unlike in the LP model, in OP we are free to choose any spatial resolution $M$ (up to the resolution of raw observations) without causing undersampling problems or high computational cost due to the effect on the state space. No free parameters are introduced by increasing the spatial resolution.

\subsubsection{Augmented Markov chain for model identifiability}
\label{subsubsec:augmentedstates}
The model described above is not identifiable because there are subsets of parameters that are exchangeable in prior distribution and which under arbitrary permutations of the state label leave the likelihood invariant (\cite{Scott2002}). This is the case for $\{\lambda_{1, n}, \lambda_{2, n}, \dots, \lambda_{\kappa, n}\}$ for each $n$ and for $\{\xi_1, \xi_2, \dots,\linebreak \xi_{\kappa}\}$. We make use of a reformulation of the model suggested by \cite{Chopin2007} to make the model identifiable, and which also readily accommodates inference for $\kappa$.

Since state labels are arbitrary, we can relabel states in order of their appearance in the Markov chain $S_{0:T}$ without affecting the model structure. This ordering of states in relation to the data means that permutations of exchangeable parameters will not leave the likelihood invariant. The relabelling is realised via the parameterisation of the Markov chain with an extension to its state space. For sequential relabelling, $s_0 = 1$, so we must have $\pi_1 = 1$ and $\pi_i = 0$ for $2 \le i \le \kappa$. We must then keep track of the number of distinct states emitted up to any time step $t$. That is, if we have $S_t = i \le K < \kappa$, we must impose the restriction that $S_{t + 1} \le K + 1$, with equality if and only if $S_{t + 1}$ has not been emitted before time $t + 1$. Thus, we let random variable $K_t$, taking values in $\{1, 2, \dots, \kappa\}$, be the number of distinct states emitted up to and including time $t$.

We can now define the augmented process $\widetilde{S}_{0:T}$ constituted by the sequence of random variables $\widetilde{S}_t \equiv \lb S_t, K_t \rb$, which have $\widetilde{\kappa} = \frac{\kappa \lb \kappa + 1 \rb}{2}$ distinct outcomes (since values are constrained by $S_t \le K_t \le \kappa$). This process is a Markov chain with transition matrix $\mathbf{\widetilde{P}} = \lb \widetilde{P}_{i, j} \rb$ for $1 \le i, j \le \widetilde{\kappa}$. If we let $i \equiv \lb s', k' \rb$, $j \equiv \lb s'', k'' \rb$, with $s', s'', k', k'' \in \{1, 2, \dots, \kappa\}$, we have
\begin{equation}
\label{eq:augmentedtransitionmat}
\widetilde{P}_{i, j}   =
\begin{cases}
P_{s', s''}   &   \text{if }   s', s''   \le   k''   =   k'   \le   \kappa,\\
\sum_{s = k' + 1}^{\kappa}   P_{s', s}   &   \text{if }   s''   =   k''   =   k'   +   1   \le   \kappa,\\
0   &   \text{else}.
\end{cases}
\end{equation}
The first case of Eq. \eqref{eq:augmentedtransitionmat} corresponds to a transition between two states previously emitted. The second to emitting a new state: since states are mutually exclusive outcomes of $S_t$ the probability of transitioning from some state $s'$ to any of the previously unseen states is the sum of the transition probabilities from $s'$ to each unseen state. The last case covers the violations of the above constraints.

Observations $X_t$ and $\mathbf{Y}_t$ are considered conditionally independent of $K_t$ given $S_t$ for $1 \le t \le T$, so this reparameterisation does not alter the dependence structure between state and observation variables of Fig. \ref{fig:dag}.

\subsection{Priors and full conditionals}
\label{subsec:priorsandfullconditionals}
This section describes prior distributions and full conditional distributions for the model parameters. These are required for the posterior sampling of parameters as part of the SMC algorithm for Bayesian parameter inference and model selection, explained in Section \ref{subsubsec:smc}.

We assume a hierarchical model structure with the following factorisation for the prior of $\mathbf{\theta}$ and $\kappa$: 
\begin{equation}
\label{eq:priorfactorisation}
p \lb   \mathbf{\theta},   \kappa   \mid   \mathbf{\phi}   \rb
=   p \lb   \mathbf{\theta}   \mid   \kappa,   \mathbf{\phi}   \rb   p \lb   \kappa   \mid   \mathbf{\phi}   \rb,
\end{equation}
in which $\mathbf{\phi}$ is the set of all hyperparameters. This allows us to efficiently sample $\lb \mathbf{\theta}, \kappa \rb$ by first sampling $\kappa$. This task is facilitated by assuming that model parameters in $\mathbf{\theta}$, with $\mathbf{P}$ considered as $\kappa$ row vectors $\mathbf{P}_{i, \cdot}$, are conditionally independent of each other given $\kappa$ and $\mathbf{\phi}$. This gives us the factorisation
\begin{equation}
p \lb   \mathbf{\theta}   \mid   \kappa,   \mathbf{\phi}   \rb
=   p \lb   \mathbf{\pi}   \mid   \kappa,   \mathbf{\phi}   \rb
\prod_{i = 1}^{\kappa}   p \lb   \mathbf{P}_{i, \cdot}   \mid   \kappa,   \mathbf{\phi}   \rb
p \lb   \xi_i   \mid   \kappa,   \mathbf{\phi}   \rb
p \lb   \Sigma_i   \mid   \kappa,   \mathbf{\phi}   \rb
\prod_{n = 1}^C   p \lb   \lambda_{i, n}   \mid   \kappa,   \mathbf{\phi}   \rb,
\end{equation}
and thus we may sample each parameter from its respective marginal prior independently, conditional on a value for $\kappa$. For each marginal prior we use a distribution conjugate to the relevant likelihood function, to facilitate sampling using standard distributions, and we fix all hyperparameters with constant values that give rise to uninformative priors.

For $\kappa$, we assume a discrete uniform prior with parameter $\bar{\kappa} \in \mathbf{\phi}$, a positive integer. That is, $\kappa$ can take on values a priori at random between $1$ and $\bar{\kappa}$. We must choose $\bar{\kappa}$ to be great enough that all model sizes that may be appropriate to the data are possible, but we are subject to increasing computational costs with larger $\bar{\kappa}$. Appropriate values can be arrived at by initial exploratory runs of the algorithm in Section \ref{subsubsec:smc}.

Priors for each parameter in $\theta$ are described in the remainder of this section along with a discussion of the corresponding full conditionals, $p \lb   \vartheta   \mid   x_{1:t},   \mathbf{y}_{1:t},   s_{0:t},   \mathbf{\theta} \setminus \vartheta,   \kappa,   \mathbf{\phi}   \rb$ for some variable $\vartheta \in \theta$, restricted to time $t$. Note we are not required to sample parameters of the initial state distribution $\mathbf{\pi}$ because the initial state is fixed at $1$ (cf. Section \ref{subsubsec:augmentedstates}).

\paragraph{Firing rates}
For the mean firing rates $\lambda_{i, n}$ we take a Gamma prior $\texttt{Gam}(\lambda_{i, n}; \alpha, \beta)$, with shape parameter $\alpha$ and rate parameter $\beta$, which is the conjugate prior for these parameters. Values of $\alpha = \frac{1}{2}, \beta = 0$ correspond to the uninformative Jeffreys prior (\cite{Gelman2003}, p69). This prior is improper and cannot be sampled from, so we use $\beta = 0.01$ for a practical alternative that is largely uninformative.

The full conditional distribution for $\lambda_{i, n}$ at time step $t$ is $\texttt{Gam}(\lambda_{i, n}; \alpha^*, \beta^*)$ with
\begin{align}
\label{eq:posteriorgamma}
\alpha^*   =   &   \sum_{\substack{u \le t: s_u = i}}   y_{t, n}   +   \alpha,\\
\beta^*   =   &   \delta t   c_{i, t}   +   \beta,
\end{align}
where $c_{i, t}   \defeq   \#   \{   s_u = i   \}_{u = 1} ^ t$; see Appendix \ref{subapp:posteriorlambda} for derivation.

\paragraph{Position model modes}
For the position hyperparameter $\xi_i$ we use as prior the discrete uniform distribution over positions $1$ to $M$. Note that we could consider $\xi_i$ as the mean of a Gaussian distribution, for which a Gaussian distribution is the conjugate prior, but for sampling from an uninformative prior with our discretisation of positions the uniform distribution is equivalent and simpler.

The full conditional distribution has the same form as the likelihood, since
\begin{align}
\label{eq:posteriorofmean1}
p \lb   \xi_i   \mid   x_{1:t},   \mathbf{y}_{1:t},   s_{0:t},   \mathbf{\theta},   \kappa,   \mathbf{\phi}   \rb
\propto   &   p \lb   x_{1:t}   \mid   s_{0:t},   \mathbf{\theta},   \kappa,   \rb
p \lb   \xi_i   \mid   \mathbf{\phi},   \kappa   \rb   \nonumber\\
\propto   &   p \lb   x_{1:t}   \mid   s_{0:t},   \mathbf{\theta},   \kappa,   \rb   \nonumber\\
\propto   &   \prod_{\substack{u \le t: s_u = i}}   p \lb   x_u   \mid   i,   \xi_i,   \Sigma_i   \rb,
\end{align}
and furthermore
\begin{equation}
\label{eq:posteriorofmean2}
p \lb   x_u   \mid   i,   \xi_i,   \Sigma_i   \rb
\propto   q \lb   \mathbf{f}_{\xi_i} \lb   x_u   \rb;   0,   \Sigma_i   \rb
\end{equation}
by Eq. \eqref{eq:positionprob}, so the posterior is $\texttt{N} \lb   \mathbf{f}_{\xi^*} \lb   \xi_i   \rb;   0,   \Sigma^*   \rb$ with
\begin{align}
\label{eq:posteriorgaussian}
\xi^*   =   &   \bar{x}_i   \in   \argmin_{x   \in   \{1, 2, \dots, M\}   }   \lcb c_{i, t}^{-1}   \sum_{\substack{u \le t: s_u = i}}   \mathbf{f}_x \lb   x_u   \rb \rcb,\\
\Sigma^*   =   &   c_{i, t}^{-1}   \Sigma_i,
\end{align}
which is derived in Appendix \ref{subapp:posteriormean}. Via this construction we can sample $\xi_i$ from the categorical distribution with probabilities obtained from $\texttt{N} \lb   \mathbf{f}_{\xi^*} \lb   \xi_i   \rb;   0,   \Sigma^*   \rb$ and normalised as in Eq. \eqref{eq:positionprob}.

\paragraph{Position model covariance matrices}
We use the conjugate Inverse-Wishart distribution as prior for $\Sigma_i$, with parameters $\Psi$ and $\delta$. This prior expresses our conception of how states characterise variability in size and shape of the regions represented in our model. These regions can be likened to place fields but for a population of place cells: they emerge from the collective activity of multiple cells. This interpretation may guide our parameterisation of this prior, since it is difficult to specify an uninformative prior over covariance matrices. The hyperparameter $\Psi$ is the $2 \times 2$ positive definite matrix of sums of squared deviations of positions transformed by $\mathbf{f}_{\xi_i}$, a priori, and $\delta$ is the degrees of freedom of the data from which $\Psi$ was derived. Thus, $\Psi$ can be set to encode our indifference to orientation or skewness of regions represented by each state by putting $\Psi_{1, 1} = \Psi_{2, 2}$ and $\Psi_{1, 2} = \Psi_{2, 1} = 0$. This leaves $\Psi_{1, 1}$ free, to be set according to our prior conception of how large these regions typically are. The influence of this hyperparameter on the prior is weighted by $\delta$; therefore a relatively uninformative prior is achieved by setting $\delta$ small (relative to the number of time bins in the data set). The full conditional for $\Sigma_i$, also Inverse-Wishart by the conjugate relationship to the Gaussian likelihood with known mean, has parameters (\cite{Gelman2003}, p87)
\begin{align}
\Psi^*   =   &   \Psi   +   SS_{i, t} \lb   \xi_i   \rb\\
\delta^*   =   &   \delta   +   c_{i, t},
\end{align}
where $SS_{i, t} \lb   \xi_i   \rb$ is the $2\times 2$ matrix of sums of squared deviations around $\xi_i$ in the transformed space,
\begin{equation}
SS_{i, t} \lb   \xi_i   \rb   \defeq   \sum_{\substack{u \le t: s_u = i}}   \mathbf{f}_{\xi_i} \lb   x_u   \rb^\intercal   \mathbf{f}_{\xi_i} \lb   x_u   \rb.
\end{equation}
Note that in the full conditionals for $\xi_i$ or $\Sigma_i$, the other parameter is considered known. In sampling procedures, we therefore either sample $\xi_i$ first conditional upon the value of $\Sigma_i$ previously sampled, or vice versa.

\paragraph{Rows of the transition matrix}
We use the Dirichlet prior for rows of $\mathbf{P}$; that is, $\texttt{Dir}(\mathbf{P}_{i, \cdot}; \mathbf{\omega})$. For an uninformative prior, we use a vector of $\kappa$ ones for $\mathbf{\omega}$.

The structure we imposed on $\mathbf{P}$ (cf. Section \ref{subsubsec:augmentedstates}) means the full conditional for a row $\mathbf{P}_{i, \cdot}$ is a \emph{Generalised Dirichlet distribution} rather than a standard Dirichlet distribution. At time step $t$ this is derived as
\begin{align}
\label{eq:posteriortransitionmatfactorisation}
&   p \lb   \mathbf{P}_{i, \cdot}   \mid   x_{1:t},   \mathbf{y}_{1:t},   \widetilde{s}_{0:t},   \mathbf{\theta},   \kappa,   \mathbf{\phi}   \rb   \nonumber\\
&\quad\propto   p \lb   s_{1:t}   \mid   k_{1:t},   \mathbf{\omega}   \rb
p \lb   \mathbf{P}_{i, \cdot}   \mid   \mathbf{\omega},   \kappa   \rb   \nonumber\\
&\quad\propto   \prod_{\substack{u \le t: s_{u - 1} = i,\\ k_u = k_{u - 1}}}   p \lb   S_u = s_u   \mid   S_{u - 1} = i,   \mathbf{P}_{i, \cdot}   \rb
\prod_{\substack{u \le t: s_{u - 1} = i,\\ k_u = k_{u - 1} + 1}}   p \lb   S_u = s_u   \mid   S_{u - 1} = i,   \mathbf{P}_{i, \cdot}   \rb   \nonumber\\
&\qquad \times   p \lb   \mathbf{P}_{i, \cdot}   \mid   \mathbf{\omega},   \kappa   \rb.
\end{align}
Note we can ignore $s_0$ because $\mathbf{\pi}$ is constant. The factorisation of $p \lb   s_{1:t}   \mid   k_{1:t},   \mathbf{\omega}   \rb$ in Eq. \eqref{eq:posteriortransitionmatfactorisation} follows from the Markov property; the first factor consists of transition probabilities between states previously emitted by the Markov chain, the second consists of transition probabilities to new states. Recall from Eq. \eqref{eq:augmentedtransitionmat} that these are treated differently. Continuing Eq. \eqref{eq:posteriortransitionmatfactorisation} we have
\begin{align}
\label{eq:posteriortransitionmat}
&   p \lb   \mathbf{P}_{i, \cdot}   \mid   x_{1:t},   \mathbf{y}_{1:t},   \widetilde{s}_{0:t},   \mathbf{\theta},   \kappa,   \mathbf{\phi}   \rb   \nonumber\\
&\quad   \propto   \prod_{j = 1}^\kappa   P_{i, j}^{   A_{i, j}(t)   -   B_{i, j}(t)   }
\prod_{j = 1}^\kappa    \lb   \sum_{l = j + 1}^{\kappa}   P_{i, l}   \rb   ^   {B_{i, j}(t)}
p \lb   \mathbf{P}_{i, \cdot}   \mid   \mathbf{\omega},   \kappa   \rb   \nonumber\\
&\quad   =   \prod_{j = 1}^\kappa P_{i, j}^{   A_{i, j}(t)   -   B_{i, j}(t)   +   \omega_j   -   1   }   \lb   \sum_{l = j + 1}^{\kappa}   P_{i, l}   \rb   ^   {B_{i, j}(t)},
\end{align}
where $\mathbf{A}(t)$ is the matrix of transition counts at time step $t$,
\begin{equation}
A_{i, j}(t)   \defeq   \sum_{u = 1}^t   \mathds{1}   \{s_u = j,   s_{u - 1} = i\},
\end{equation}
and $\mathbf{B}(t)$ is the matrix of first arrival indicator variables at time step $t$,
\begin{equation}
B_{i, j}(t)   \defeq
\begin{cases}
1,   \quad   \text{the first $j$ in $s_{1:t}$ immediately follows $i$,}\\
0,   \quad   \text{else,}
\end{cases}
\end{equation}
for $1 \le i, j \le \kappa$. The posterior probabilities given by Eq. \eqref{eq:posteriortransitionmat} correspond to a Generalised Dirichlet distribution with parameters $\mathbf{\zeta}_i   =   \mathbf{A}_{i, \cdot}(t)   -   \mathbf{B}_{i, \cdot}(t)   +   \mathbf{\omega}$ and $\mathbf{\gamma}_i   =   \mathbf{B}_{i, \cdot}(t)$ (\cite{Wong1998}). We can use the algorithm of \cite{Wong1998} to efficiently sample from this posterior; details are provided in Appendix \ref{subapp:transitionmatrows}.

\subsection{Inference with our model}
\label{subsec:inferenceinmodel}
There are four kinds of inference we are interested in and can perform with our model. The first is inference for model parameters $\theta$. Section \ref{subsubsec:smc} describes the algorithm we use to estimate the posterior distribution over these parameters, and Section \ref{subsubsec:estimation} explains how we use the posterior expectation as point estimate for $\theta$. Secondly, for states $S_{0:T}$: this is explained in Section \ref{subsubsec:statedecoding}, in which is also also explained how we arrive at an estimate for $\kappa$. Thirdly, for position variables $X_{1:T}$ from spike train observations $\mathbf{Y}_{1:T}$: \emph{decoding} position, explained in Section \ref{subsubsec:positiondecoding}. The fourth kind of inference is for the occurrence of replay in REST data. The analysis of replay is treated in Section \ref{subsec:replaydetection}.

\subsubsection{Sequential Monte Carlo (SMC) algorithm for Bayesian parameter inference}
\label{subsubsec:smc}
For the inference of model parameters $\theta$ and $\kappa$ we target the posterior distribution $p \lb   \mathbf{\theta},   \kappa   \mid   x_{1:T},   \mathbf{y}_{1:T},   \mathbf{\phi}   \rb$. The necessary marginalisation of the state process $\widetilde{S}_{0:T}$ is only computationally feasible when $T$ is far smaller than what we must use in experimental data. For this reason we turn to sampling-based procedures such as Gibbs sampling, which are commonly employed in similar settings. However, as explained in \cite{Chopin2007} and explored in \cite{Celeux2000}, even when the model is identifiable and $\kappa$ is fixed, Gibbs sampling for HMM parameters can fail to mix efficiently and can spend too much time exploring uninteresting local maxima of parameter space, due to the complexity of the data.

The SMC algorithm of \cite{Chopin2007} addresses this by using importance-weighted ``particles'' to sample the \emph{partial} posterior distributions, $p \lb   \mathbf{\theta},   \kappa   \mid   x_{1:t},   \mathbf{y}_{1:t},   \mathbf{\phi}   \rb$ for $1 \le t \le T$. Since the partial posteriors when $t$ is small tend to be much flatter than the full posterior, particles are more readily able to escape inferior modes. A ``resample-move'' step effects an exploration of parameter space, and rejuvenates the sample when it becomes degenerate as new data is accumulated. An outline of the algorithm follows.

\paragraph{- Initialisation:}
Use Eq. \eqref{eq:priorfactorisation} to sample $\kappa$, and $\mathbf{\theta}$ conditional on sampled values of $\kappa$, $H$ times, obtaining $\{\mathbf{\theta}^h, \kappa^h\}_{h = 1}^H$. We refer to the set of all particles that sample the same value of $\kappa$ as the \emph{subpopulation corresponding to $\kappa$}. Initialise the \emph{particle weights} as
\begin{equation}
w_h   \gets   \frac{1}{H}   \qquad   \text{for}   \qquad   h   =   1, 2, \dots, H.
\end{equation}
\paragraph{- Loop:}
At each time step $t$ from $1$ to $T$, perform all or some of the following tasks as necessary:

\paragraph{(1) Update weights:}
Set
\begin{equation}
w_h   \gets   w_h   p \lb    x_t,   \mathbf{y}_t   \mid   x_{1:t - 1},   \mathbf{y}_{1:t - 1},   \mathbf{\theta}^h   \kappa^h   \rb
\end{equation}
for $h = 1, 2, \dots, H$. The weight update factor is the ratio of \emph{data} likelihoods at subsequent time steps:
\begin{equation}
p \lb   x_t,   \mathbf{y}_t   \mid   x_{1:t - 1},   \mathbf{y}_{1:t - 1},   \mathbf{\theta}^h,   \kappa^h   \rb
=   \frac{   p \lb   x_{1:t},   \mathbf{y}_{1:t}   \mid   \mathbf{\theta}^h,   \kappa^h   \rb   }{   p \lb   x_{1:t - 1},   \mathbf{y}_{1:t - 1}   \mid   \mathbf{\theta}^h,   \kappa^h   \rb   }
\end{equation}
(using $p \lb   x_1,   \mathbf{y}_1   \mid   \mathbf{\theta}^h,   \kappa^h   \rb$ at $t = 1$). The data likelihood at $t$ can be computed by marginalising $\widetilde{S}_t$ from the \emph{forward function at $t$}, $p \lb   \widetilde{S}_t,   x_{1:t},   \mathbf{y}_{1:t}   \mid   \mathbf{\theta}^h,   \kappa^h   \rb$, computed using the \emph{forward recursions}, explained in \cite{Scott2002}.

\paragraph{(2) Check for sample degeneracy:}
\tolerance=450
evaluate the effective sample size (ESS) (\cite{Kong1994})
\begin{equation}
ESS   =   \frac{H}{1   +   var \lb   w   \rb}
\end{equation}
using the sample variance of the weights. This being small relative to $H$ indicates that the sample is of poor quality in light of recent observations. If $ESS$ exceeds a threshold $ESS^*$, skip \emph{(3)} and \emph{(4)} and proceed to the next time step.

\paragraph{(3) Resample with positive discrimination and reset weights:}
resample particles according to their weights (using, for example, the \emph{residual resampling} approach of \cite{Liu1998}). This should be done in conjunction with the ``positive discrimination'' scheme of \cite{Chopin2007}, to make it more likely we retain some particles in each subpopulation after resampling. Compute the sample approximation to the marginal (partial) posterior over $\kappa$:
\begin{equation}
\label{eq:kappaposterior}
\hat{p}_{\kappa', t}   \defeq   p \lb   \kappa = \kappa'   \mid   x_{1:t},   \mathbf{y}_{1:t},   \mathbf{\phi}   \rb
=   \frac{   \sum_{h: \kappa^h = \kappa'}  w_h   }{\sum_{h = 1}^H   w_h   }
\end{equation}
for each $1 \le \kappa' \le \bar{\kappa}$. If $\hat{p}_{\kappa', t}H$ falls below a tolerance level $H^*$, resample $H^*$ times from the subpopulation corresponding $\kappa'$. For these resampled particles, set
\begin{equation}
\label{eq:discriminatedweights}
w_h   \gets   \frac{   \hat{p}_{\kappa', t}   H   }{   H^*   }   <   1.
\end{equation}
We thus give discriminated particles lower importance, compensating for the biasing effect of their preferential retention. \cite{Chopin2007} suggests to use $H^* = \frac{H}{10}$.

After resampling within all subpopulations requiring positive discrimination, resample the remaining particles maintaining a sample size of $H$ and set their weights to $1$, then normalise all weights.

Resampling purges the sample of particles with low importance and replenishes it with copies of particles with high importance. This focusses the attention of the sampler on promising regions of parameter space. \cite{Chopin2007} suggests a threshold of $ESS^* = \frac{H}{2}$. ``Positive discrimination'' is necessary because traditional resampling cannot refresh the $\kappa$ component of the sample because of the dependence of $\mathbf{\theta}$ on $\kappa$. Consequently, if resampling should cause one subpopulation of particles to become empty there is no mechanism for replenishing it. This is a problem if it occurs before enough observations have been taken into consideration to confidently rule on whether the corresponding value of $\kappa$ is worth exploring further, and is particularly a danger in early time steps for particles with large sampled $\kappa$, for if these should be lost during the time when $\kappa$ appears to be small, there will be no representation of large $\kappa$ later on when warranted by the further accumulation of data.

\paragraph{(4) ``Move'' particles using a single sweep of Gibbs sampling:}
for each $1 \le h \le H$, sample $\widetilde{s}_{0:t}^h$ according to the distribution $p \lb   \widetilde{s}_{0:t}   \mid   x_{1:t},   \mathbf{y}_{1:t},   \mathbf{\theta}^h,   \kappa^h   \rb$ using the \emph{stochastic backward recursions} (described in \cite{Scott2002}), then sample $\theta^h$ according to the posterior $p \lb   \mathbf{\theta}_t   \mid   s_{0:t}^h,   x_{1:t},   \mathbf{y}_{1:t},   \mathbf{\theta}^h,   \kappa^h   \rb$ described in Section \ref{subsec:priorsandfullconditionals}.

By our use of conjugate priors, we are only required to compute the statistics $\mathbf{A}(t), \mathbf{B}(t), c_{i, t}, SS\lb i, t, \xi_i \rb$ and $\bar{x}_i$ for $1 \le i \le \kappa$, then to sample from standard distributions. We perform sampling from the Gamma, Inverse-Wishart, and Beta distributions using built-in functions of software package MATLAB.

\subsubsection{Parameter estimation}
\label{subsubsec:estimation}
The algorithm of Section \ref{subsubsec:smc} results in a sample approximation to $p \lb   \mathbf{\theta},   \kappa   \mid   x_{1:T},   \mathbf{y}_{1:T},   \mathbf{\phi}   \rb$; we make a point estimate $\hat{\mathbf{\theta}}$ of $\mathbf{\theta}$ using the sample posterior mean. For a particular parameter $\vartheta_i$ associated with state $i$, we have
\begin{equation}
\hat{\vartheta}_i   =   \frac{   \sum_{h: \kappa^h \ge i}   w_h   \vartheta_i^h   }{   \sum_{h: \kappa^h \ge i}   w_h   }.
\end{equation}
This achieves a marginalisation of $\kappa$.

\subsubsection{State estimation}
\label{subsubsec:statedecoding}
We use the smoothed posterior distributions over $S_t, K_t$ to estimate the state variable at each time step and the number of states $\kappa$. Inference for $\kappa$ could be performed via Eq. \eqref{eq:kappaposterior} with an estimate $\hat{\kappa}$ taken as the mode; however, as argued in \cite{Chopin2007} this is an estimate of how many states would be observed eventually if we took enough observations and one should use the posterior distribution of $K_T$ to estimate how many distinct states were emitted during the $T$ time steps. Thus, after fixing $\theta$ to our estimates $\hat{\theta}$, we use the forward-backward algorithm to compute the smoothed posterior distributions
\begin{equation}
\label{eq:smoothingposterior}
p \lb   S_t = i,   K_t = k   \mid   x_{1:T},   \mathbf{y}_{1:T},   \hat{\theta}   \rb,
\end{equation}
for all $\lb i, k \rb \in \lcb 1, 2, \dots, \kappa \rcb ^ 2$ and for all $t \in \lcb 1, 2, \dots, T \rcb$. We obtain the marginal distribution over $K_t$ by summing Eq. \eqref{eq:smoothingposterior} over all $\bar{\kappa}$ values of $S_t$, and vice versa for $S_t$. The maximum a posteriori (MAP) estimate at time step $t$ is the value that maximises the marginal posterior distribution. We take the MAP estimate of $K_T$ for $\hat{\kappa}$, our estimate of the number of states required to characterise the data. We can alternatively use the Viterbi algorithm, described in \cite{Scott2002}, which returns the sequence $\widetilde{s}_{0:T}$ of greatest posterior probability, i.e. the sequence that maximises $p \lb   \widetilde{s}_{0:T}   \mid   x_{1:T}, \mathbf{y}_{1:T}, \hat{\theta}   \rb$.

\subsubsection{Position decoding}
\label{subsubsec:positiondecoding}
We can also use our model to estimate (decode) position at any time from spike train observations. We can compute the position posterior distributions, $p \lb   x_t   \mid   \mathbf{y}_{1:T}, \hat{\theta}   \rb$, and hence obtain the MAP point estimate, as used by other authors in studies of replay such as \cite{Davidson2009}. To do this we take advantage of the conditional independence of $X_t$ from $\mathbf{Y}_{1:T}$ given $S_t$, which permits
\begin{equation}
\label{eq:positionposterior}
p \lb   X_t = x, S_t = i   \mid   \mathbf{y}_{1:T}, \hat{\theta}   \rb
=   p \lb   S_t = i   \mid   \mathbf{y}_{1:T},   \hat{\theta}   \rb
p \lb   X_t = x   \mid   S_t = i,   \hat{\xi}_i,   \hat{\Sigma}_i   \rb.
\end{equation}
On the right hand side of Eq. \eqref{eq:positionposterior} is the marginal smoothing posterior at time step $t$ using spike train observations only, and the conditional probability over positions given state, using the fitted model parameters. We then obtain the position posterior distribution by marginalising $S_t$.

We can instead compute the trajectory $\hat{x}_{1:T}$ of greatest posterior probability, i.e. that maximises $p \lb   x_{1:T}   \mid   \mathbf{y}_{1:T}, \hat{\theta}   \rb$. For this we use a modified version of the Viterbi algorithm, explained in Appendix \ref{app:viterbiforposition}.

\subsection{Model-based replay detection}
\label{subsec:replaydetection}
In an analysis of sleep replay we wish to make three kinds of inference: the time of replay occurring, the information content being replayed, and the rate of time compression relative to the behavioural timescale. The methods described in this section allow us to achieve each of these.

Our idea is to use the posterior distribution over trajectories given spike train observations as a representation of what information is encoded at different times. We identify replay as occurring at a particular time when the posterior probability of a certain trajectory obtains a maximum above some threshold (see Section \ref{subsubsec:replayscore}). For inference regarding the information content being replayed, we fix the trajectories to be used for this posterior evaluation. We call these \emph{template} trajectories (Section \ref{subsubsec:templates}). For the rate of temporal compression, we search for replay in temporally compressed data at many different compression rates (Section \ref{subsubsec:timecompression}).

Spike train data for replay analysis may be distinct from the training data (for example when using a REST epoch for replay analysis) and therefore constitute dynamics and correlations that may not be described accurately by the model with $\mathbf{\theta} = \hat{\mathbf{\theta}}$ estimated from RUN. We must therefore demonstrate predictive power for our model with parameterisation $\hat{\mathbf{\theta}}$ on the data $\mathbf{y}^{\text{REST}}_{1:T}$, for which we use a likelihood-based method, explained in Section \ref{subsubsec:modelcomparison}.

\subsubsection{Replay score}
\label{subsubsec:replayscore}
We define the \emph{replay score}, $\Omega$, for template trajectory $x_{1:a}$ at time $t$, as the ratio of likelihoods
\begin{equation}
\label{eq:replayscore}
\Omega \lb   x_{1:a},   t;   \mathbf{y}_{1:T},   \mathbf{\theta}   \rb   \defeq   \frac{   p \lb   X_t = x_1,   \dots,   X_{t + a - 1} = x_a   \mid   \mathbf{y}_{1:T},   \mathbf{\theta}   \rb   }{   p \lb   X_t = x_1,   \dots,   X_{t + a - 1} = x_a   \mid   \mathbf{\theta}   \rb   }.
\end{equation}
An algorithm for computing the numerator of Eq. \eqref{eq:replayscore} is described in Appendix \ref{app:trajectoryprobability}, and for the denominator in Appendix \ref{app:marginaltrajectoryprobability}. Then we say that template $x_{1:a}$ is \emph{replayed at time $t^{rep}$}, on the discrete timescale, if
\begin{equation}
\label{eq:replaythreshold1}
\Omega = \Omega \lb   x_{1:a},   t^{rep};   \mathbf{y}^{\text{REST}}_{1:T},   \hat{\mathbf{\theta}}   \rb   >   \Omega^*
\end{equation}
and
\begin{equation}
\label{eq:replaythreshold2}
\Omega   >   \max \lcb   \Omega \lb   x_{1:a},   t^{rep} - 1;   \mathbf{y}^{\text{REST}}_{1:T},   \hat{\mathbf{\theta}}   \rb   ,
\Omega \lb   x_{1:a},   t^{rep} + 1;   \mathbf{y}^{\text{REST}}_{1:T},   \hat{\mathbf{\theta}}   \rb\rcb,
\end{equation}
for some threshold $\Omega^*$, where $\hat{\mathbf{\theta}}$ are the model parameters estimated from RUN. Since Eq. \eqref{eq:replayscore} has the form of a model likelihood ratio between the model for trajectories conditional on spike train observations and the model for trajectories marginal of spike trains, in our applications we use for $\Omega^*$ values suggested by \cite{Kass1995} for likelihood ratios in Bayesian model comparison. Those authors provide useful interpretations for this ratio, in particular that $\Omega^* = 20$ is the minimum for ``strong'' evidence and $\Omega^* = 150$ for ``very strong'' evidence.

\subsubsection{Templates}
\label{subsubsec:templates}
\begin{figure}[h]
\centering
\includegraphics[scale=0.2]{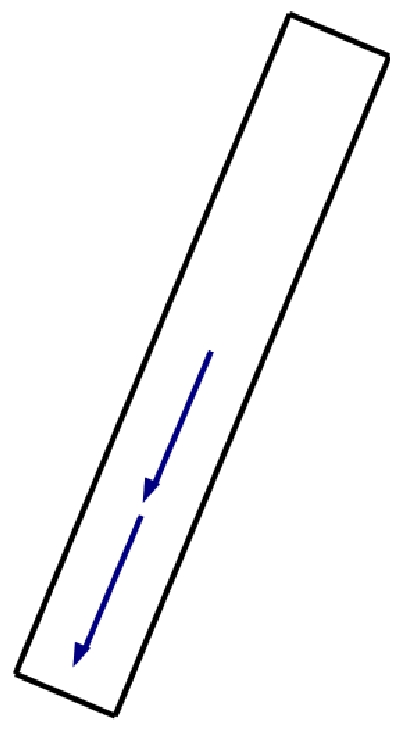}
\hspace{1.0cm}
\includegraphics[scale=0.2]{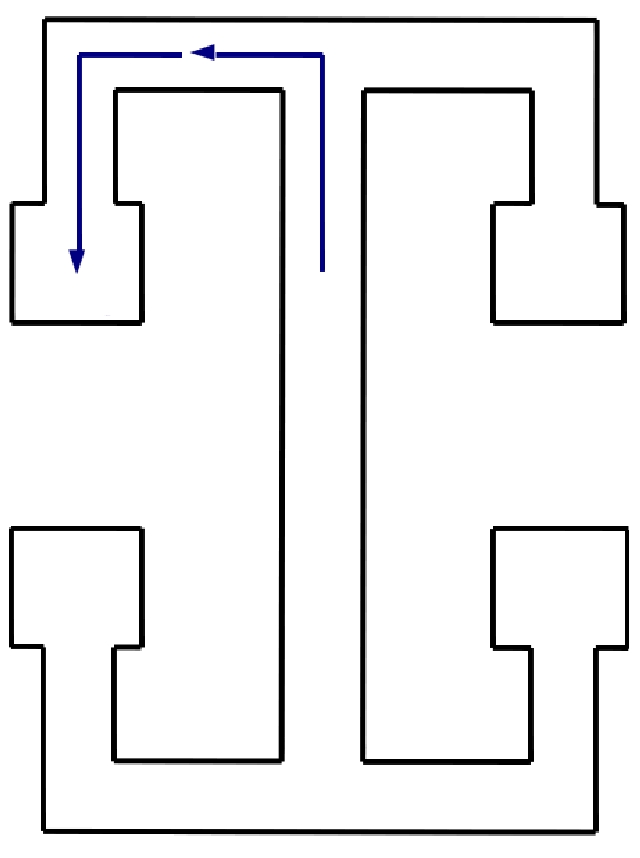}
\caption{Top-down outline of the two environments used in RUN data (not to scale). Blue arrows represent example template trajectories used for replay detection. \emph{Left:} linear track. \emph{Right:} T-maze.}
\label{fig:templates}
\end{figure}

We describe a collection of trajectories of the form $x_{1:a}$ to use in Eq. \eqref{eq:replayscore}. For the results presented in Section \ref{subsec:replayanalysisresults} we use segments of the RUN trajectory through particular regions of the environment; for example around a corner or into a rest site (on the T-maze). We chose segments running in both directions, i.e. towards and away from the centre of the environment. Examples of how these template trajectories might look are given in Fig. \ref{fig:templates}.

\subsubsection{Time compression}
\label{subsubsec:timecompression}
By choosing templates that represent trajectories at uncompressed (behavioural) speeds, we are able to use our replay detection method for studying replay on a rapid (compressed) time scale relative to the behavioural time scale by adjusting the time discretisation bin width used for the analysis data. That is, for the detection of replay of a template $x_{1:a}$ at compression rate $c$, we compute $\Omega$ using Eq. \eqref{eq:replayscore} on compressed spike train data $\overline{\mathbf{y}}_{1:cT}$ obtained by re-binning the raw spike train data, using the procedure of Section \ref{subsubsec:discretisation}, with bin width $\overline{\delta t} = \delta t / c$.

\subsubsection{Assessing model fit on analysis data}
\label{subsubsec:modelcomparison}
In order to justify our use of $\hat{\mathbf{\theta}}$ in Eq. \eqref{eq:replayscore}, i.e. our model fitted to a RUN data set being used for replay detection on a REST data set, we make an assessment of model fit using the data likelihood (\cite{Gelman2003}), $p \lb   \mathbf{y}^{\text{REST}}_{1:T}   \mid   \mathbf{\theta},   \kappa   \rb$. In particular we use the Bayesian information criterion (BIC, \cite{Schwarz1978})
\begin{equation}
\label{eq:bic}
BIC   =   -2   \log p \lb   \mathbf{y}^{\text{REST}}_{1:T}   \mid   \mathbf{\theta},   \kappa   \rb   +   N   \log T,
\end{equation}
where $N$ is the number of free parameters in the model ($N = \hat{\kappa} \lb \hat{\kappa} + C + 3 \rb$ for OP). A lower BIC implies a better fit to the data, and includes a penalty for larger models. We compute the BIC for various parameterisations of the model: our estimates obtained from training (RUN) data, $\hat{\theta}$, and several alternatives chosen as benchmarks for particular aspects of model fit. Firstly, the model fitted to the analysis data itself, i.e. $\theta$ is estimated from REST spike train data using the procedure of Section \ref{subsubsec:smc}, ignoring the position model. We expect the BIC for $\hat{\theta}$ estimated from RUN to be greater than this alternative, but if it is close relative to an inferior benchmark we will have evidence that $\hat{\theta}$ estimated from RUN is well fit to REST. Secondly, as an inferior benchmark, we compute the BIC for a sample of $\theta$ drawn from the prior (cf. Section \ref{subsec:priorsandfullconditionals}) and the prior mean BIC. Thirdly, the model with parameterisation $\hat{\theta}$ except for the transition matrix $\mathbf{P}$; instead we assume that $S_t$ comes from the stationary distribution (computed from $\mathbf{P}$) at each $t$. This we use to assess whether the Markovian dynamics found for the training data are beneficial to the description of the analysis data. If this alternative has a lower BIC, it suggests the dynamics described by $\mathbf{P}$, as estimated from the RUN data, do not also describe the REST data as well as simply assuming independence through time. Fourthly, BD, as described by \cite{Zhang1998} and with parameters estimated from RUN using maximum likelihood.

\subsubsection{Replay detection algorithm}
\label{subsubsec:replaydetectionalgorithm}
We can now state our replay detection algorithm as follows:
\begin{enumerate}
\item Use training data (a RUN epoch) and the procedure of Sections \ref{subsubsec:smc} and \ref{subsubsec:estimation} to estimate model parameters as $\hat{\mathbf{\theta}}$.
\item Use the model comparison approach of Section \ref{subsubsec:modelcomparison} to verify the fitted model can be used on the analysis (REST) data.
\item Construct a set of templates $\{x_{1:a_r}^{(r)}\}_{r = 1}^R$.
\item Evaluate Eq. \ref{eq:replayscore} for each template $x_{1:a_r}^{(r)}$ and for $t = 1, \dots, T - a_r + 1$.
\item Report $t^{rep}$ as a replay of template $r$ whenever Eqs. \eqref{eq:replaythreshold1} and \eqref{eq:replaythreshold2} are satisfied at $t^{rep}$ for $x_{1:a_r}^{(r)}$.
\end{enumerate}

Times of replay events detected using this procedure at different compression rates are then classified as distinct events only when the extent of their temporal overlap is less than $50\%$. This is necessary because the time of the event, as indicated by a local optimum of $\Omega$, is liable to change between compression rates since slight adjustments to the placement of the template may improve the score. This rule is applied also to events detected using different templates: when two or more detected events overlapped by at least $50\%$, the event with greatest $\Omega$ was retained and all others discarded, to prevent multiple discoveries of the same event.

\subsection{Correlation of replay with SWR events}
\label{subsec:replayripplecorrelation}
We use the cross-correlogram between replay events and SWR events to demonstrate correlation between these two processes. SWR events were detected by bandpass filtering LFP between $120$Hz and $250$Hz, then taking the times of peak filtered LFP during intervals exceeding $3.5$ standard deviations. In addition, we required that these intervals were between $30$ms and $500$ms in duration, between $20\mu$V and $800\mu$V in amplitude and with a gap between distinct intervals of at least $50$ms.

The correlation between the process consisting of replay events (\emph{rep}) and the process of SWR events (\emph{rip}) at a temporal offset $u$ seconds from any time $t$ is measured by the second-order product density function for stationary point processes (\cite{Brillinger1976}),
\begin{equation}
\label{eq:crossproductdensity}
\rho_{rep, rip} \lb   u   \rb
\defeq   \lim_{h, h' \to 0}   Pr \lb   rep \text{ event in } (t + u, t + u + h],
rip \text{ event in } (t, t + h']   \rb   /   h h'.
\end{equation}
An unbiased estimator of this is
\begin{equation}
\label{eq:crossproductdensityestimator}
\hat{\rho}_{rep, rip} \lb   u   \rb   =   \lb   \tau T \delta t   \rb ^ {-1}   J_{rep, rip} \lb   u   \rb
\end{equation}
(\cite{Brillinger1976}), in which $J_{rep, rip} \lb   u   \rb$ is the cross correlogram at lag $u$ with bin width $\tau$,
\begin{equation}
\label{eq:crosscorrelogram}
J_{rep, rip} \lb   u   \rb   \defeq   \#  \lcb   \lb i, j \rb   :   u - \tau / 2  <   t^{rep}_i - t^{rip}_j   <   u + \tau / 2,
t^{rep}_i \ne t^{rip}_j   \rcb,
\end{equation}
where $t^{rep}_i, t^{rip}_j$ are times of replay events and SWR events respectively (thus, for positive intervals $t^{rep}_i - t^{rip}_j$ the SWR event occurs first), and $T \delta t$ is the observed duration of the two processes, in seconds. The discretisation parameter $\delta t$ of our model and the average duration of SWR events determine the minimum discernable lag between replay and SWR events, and thus our choice of $\tau$.

We compare $\hat{\rho}_{rep, rip} \lb   u   \rb$ at various lags $u$ with the theoretical value of Eq. \eqref{eq:crossproductdensity} for unrelated processes, estimated by $N_{rep} \lb   T \delta t   \rb   N_{rip} \lb   T \delta t   \rb   /   \lb T \delta t \rb ^ 2$, where $N_a \lb   t   \rb$ is the number of events of point process $a$ in the interval $(0, t]$. $\hat{\rho}_{rep, rip} \lb   u   \rb$ being greater than this for lags close to zero signifies that events of the processes occur at approximately the same time.

\cite{Brillinger1976} shows that, for $T \delta t \to \infty$, for each $u$ separated by $\tau$, the $J_{rep, rip} \lb   u   \rb$ follow independent Poisson distributions with parameter $T \delta t \tau \rho_{rep, rip} \lb   u   \rb$. The dependence of the estimator distribution on the parameter being estimated suggests a variance-stabilising square root transformation. Thus, independently for each $u$, $\sqrt{   \hat{\rho}_{rep, rip} \lb   u   \rb   }$ is approximately distributed as $\texttt{N} \lb   \sqrt{ \rho_{rep, rip} \lb   u   \rb },   \lb   4 T \delta t \tau    \rb ^ {-1}   \rb$. We use this fact to construct $\lb 1 - \alpha \rb \%$ confidence intervals around the estimates. We adjust the significance level $\alpha$ to account for our making multiple comparisons (one at each lag $u$) using the Bonferroni correction, which is to divide $\alpha$ by the number of comparisons made. This is very conservative as we are only interested in lags close to zero.

\subsection{Data simulation}
\label{subsec:simulation}
We used simulated data (spike trains and position trajectory) to evaluate our parameter inference algorithm and our replay detection algorithm. The general simulation method, in which the parameterisation $\theta, \kappa$ is prespecified and data randomly simulated from the model with this parameterisation, is explained in Section \ref{subsubsec:datasimulation}. Section \ref{subsubsec:replaysimulation} explains how we simulate a set of spike trains in which multiple instances of a trajectory segment are encoded for the purpose of evaluating our replay detection algorithm.

\subsubsection{Simulation of observation processes}
\label{subsubsec:datasimulation}
For the evaluation of our parameter inference algorithm, we used a known parameterisation of the model to simulate spike trains and positions from we which made estimates of the parameters to compare with the known values. We first specified a model size $\kappa^*$, then used an initial run of the algorithm of Section \ref{subsubsec:smc} with fixed state space dimension $\kappa^*$ on the experiment data to find a set of realistic parameter values $\theta^*$. Then we sampled a sequence $s_{0:T}$ by setting $s_0$ to $1$ (an arbitrary choice), then sampling $s_t$ from the discrete distribution $\mathbf{P}^*_{s_{t - 1}, \cdot}$ for $t \in \lcb 1, 2, \dots, T \rcb$. Positions and spike trains were then generated, on the discrete time scale, by sampling $x_t$ from the distribution with probabilities $p \lb X_t = x \mid S = s_t, \theta^* \rb$, and $y_{t, n}$ from $\texttt{Poi} \lb \lambda^*_{s_t, n} \rb$ for $n \in \lcb 1, 2, \dots, C \rcb$.

\subsubsection{Replay simulation}
\label{subsubsec:replaysimulation}
We assessed our replay detection algorithm of Section \ref{subsec:replaydetection} by applying it to simulated spike train data in which known replay events were inserted. Our approach was to generate spike trains that correlate (via our model) with a random hidden position trajectory punctuated by instances of the template trajectories discussed in Section \ref{subsubsec:templates}.

To achieve this we fixed $\theta^*, \kappa^*$ as in Section \ref{subsubsec:datasimulation} and simulated a full trajectory $x_{1:T}$. Then, for each of several templates $x_{1:a_r}^{(r)}$, we selected uniformly at random $N_r$ time bins between 1 and $T - a_r + 1$ as the replay event times, and at each event time $u$, we set $x_{u:u + a_r - 1} \gets x_{1:a_r}^{(r)}$. No two events were permitted to overlap: we resampled the later event time whenever this occurred. We then used the forward-backward algorithm to compute the smoothing posteriors for the state process $S_{1:T}$ using the position trajectory alone, and used these to compute the posterior mean firing rate
\begin{equation}
\label{eq:posteriormeanfiringrate}
\bar{\lambda}_n   =   \sum_{i = 1} ^ {\kappa^*}   \lambda_{i, n}   p \lb   S_t = i   \mid   x_{1:T},   \theta^*   \rb
\end{equation}
for each cell $n$, at each time step $t$, then sampled a number of spikes for cell $n$ in time bin $t$ according to the Poisson distribution with mean $\bar{\lambda}_n$.

%% file: results.tex
\section{Results}
\label{sec:results}
\subsection{Parameter and model size estimation}
\label{subsec:fittingresults}
\subsubsection{Simulated data}
\label{subsubsec:simulatedparameters}
Using the method of Section \ref{subsubsec:datasimulation}, we simulated two data sets, distinguished by the domain used for position variables: one each corresponding to the linear track environment and the T-maze. In the simulated linear track data we used $C = 4$ and $\kappa^* = 4$, and in the simulated T-maze data we used $C = 10$ and $\kappa^* = 5$. This data we supplied to our model fitting algorithm to obtain estimates $\hat{\theta}, \hat{\kappa}$.

Our algorithm correctly identified $\kappa^*$ in both data sets, using the modal value of $K_T$ as explained in Section \ref{subsubsec:statedecoding}. In Tables \ref{tab:simulatedlineartrack} and \ref{tab:simulatedtmaze} (corresponding to the linear track data and the T-maze data respectively) are measures of accuracy for our estimates of the conditional distributions over position given state and for rows of the transition matrix, by means of the Kullback-Leibler (K-L) divergence from a target distribution to the estimated distribution. The K-L divergence (cf. \cite{Dayan2001}, p323) is a nonsymmetric distance between distributions; it has a minimum of 0, which is obtained if and only if the distributions are identical. In these tables we compare the K-L divergence from each target distribution to our estimates, against the K-L divergence from the target to a uniform distribution on the same support. The uniform distribution represents an estimate based on no data. We find that the K-L divergences from the targets to our estimates is one or two orders of magnitude smaller than those to the uniform distribution for each position model, and four or more orders of magnitude smaller for each row of the transition matrix, suggesting good accuracy for our estimates.

\begin{table}[h]
\caption{Performance of model fitting algorithm: K-L divergence (in bits) of estimated model distributions, conditional on state, from target (simulated) distributions. Divergences of uniform distributions of appropriate size are provided for comparison. \emph{Data set 1: simulated linear track.}}
\label{tab:simulatedlineartrack}
\centering
\tiny
\begin{tabular*}{0.45\textwidth}{r|llll}
\hline
State & \multicolumn{2}{l}{Position model} & \multicolumn{2}{l}{Transition matrix row}\\
& Estimated & (Uniform) & Estimated & (Uniform)\\
\hline
1 & $5.39\times 10^{-2}$ & $1.78$ & $8.17\times 10^{-5}$ & $1.95$\\
2 & $1.54\times 10^{-2}$ & $1.48$ & $4.34\times 10^{-4}$ & $1.75$\\
3 & $1.16\times 10^{-1}$ & $1.39$ & $4.55\times 10^{-4}$ & $1.77$\\
4 & $1.57\times 10^{-1}$ & $1.82$ & $5.51\times 10^{-4}$ & $1.95$\\
\hline
\end{tabular*}
\end{table}
\begin{table}[h]
\caption{Kullback-Leibler divergences (in bits) of estimated model distributions from true values, as in Table \ref{tab:simulatedlineartrack}. \emph{Data set 2: simulated T-maze.}}
\label{tab:simulatedtmaze}
\centering
\tiny
\begin{tabular*}{0.45\textwidth}{r|llll}
\hline
State & \multicolumn{2}{l}{Position model} & \multicolumn{2}{l}{Transition matrix row}\\
& Estimated & (Uniform) & Estimated & (Uniform)\\
\hline
1 & $3.80\times 10^{-1}$ & $3.18$ & $9.74\times 10^{-4}$ & $2.29$\\
2 & $1.96\times 10^{-1}$ & $2.17$ & $3.89\times 10^{-3}$ & $2.26$\\
3 & $2.73\times 10^{-1}$ & $2.02$ & $9.67\times 10^{-4}$ & $2.29$\\
4 & $2.86\times 10^{-1}$ & $1.83$ & $1.62\times 10^{-3}$ & $2.28$\\
5 & $1.01\times 10^{-1}$ & $2.55$ & $2.90\times 10^{-3}$ & $2.29$\\
\hline
\end{tabular*}
\end{table}

\subsubsection{Experimental data}
\label{subsubsec:experimentalparameters}
\begin{figure*}[t]
\centering
\includegraphics[scale=0.3,natwidth=1905,natheight=498]{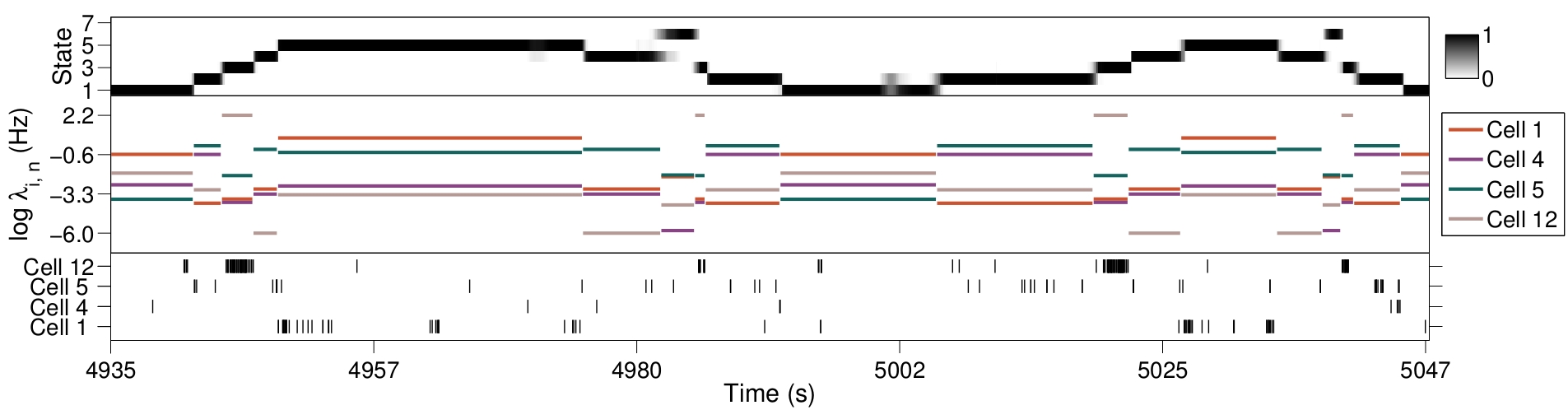}
\caption{Segment of the T-maze RUN data exhibiting the model characterisation of spike trains. \emph{Top panel:} smoothing posterior distribution over hidden state at each time step. \emph{Middle panel:} mean spike rate (in log domain for clarity) conditional on the MAP hidden state for four cells in the sample. \emph{Bottom panel:} rasters of observed spike times.}
\label{fig:fittedspiketrainparameters}
\end{figure*}

\begin{figure*}[t]
\centering
\begin{subfigure}[t]{0.45\textwidth}
\includegraphics[scale=0.42,natwidth=659,natheight=451]{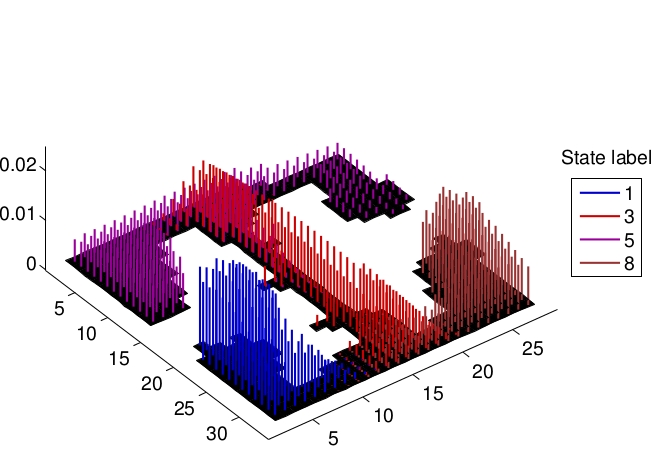}
\end{subfigure}
\hspace{0.5cm}
\begin{subfigure}[t]{0.45\textwidth}
\includegraphics[scale=0.42,natwidth=659,natheight=451]{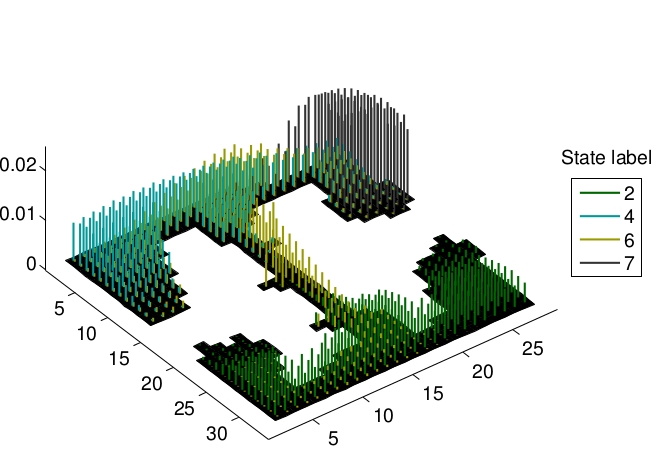}
\end{subfigure}
\caption{Model characterisation of position in T-maze data: each cluster of vertical bars of a single colour represents the distribution over the discrete positions on which they stand, conditional upon a particular state. The height of each bar represents probability mass.}
\label{fig:fittedpositiondistributions}
\end{figure*}

We applied the algorithm of Section \ref{subsec:inferenceinmodel} to the linear track and the T-maze data sets with a discretisation bin width of $\delta t = 100$ms, and found $\hat{\kappa} = 5$ for the linear track and $\hat{\kappa} = 8$ for the T-maze. For this we used $\bar{\kappa} = 10$ (after some exploratory runs of the algorithm with greater $\bar{\kappa}$ to eliminate larger models and greater $\delta t$ for faster computation) and a sample size of $H = 1500$ particles.

Fig. \ref{fig:fittedspiketrainparameters} depicts, for an interval of T-maze RUN data, the smoothing posteriors over the hidden states $S_t$ and how the changing state corresponds to changing levels of activity in the spike trains. The middle panel of the figure shows, for several cells $n$, the value of $\log \hat{\lambda}_{\hat{s}_t, n}$, with $\hat{s}_t$ the MAP state at time $t$, as a piecewise continuous line. By comparing these jumping spike rates to the spike trains represented by the raster plot in the bottom panel, one can see how different states correspond to different levels of cell activity and how the Markov chain characterises variability in the activity of all cells simultaneously.

Fig. \ref{fig:fittedpositiondistributions} depicts the estimated distributions over positions conditional on state for the T-maze data. Probabilities are represented by the height of bars and states are distinguished with different colours. These demonstrate how the states of the Markov chain constitute a coarse-grained representation of position: broad regions of the environment are associated with a particular state, characterised by a central position and covariance structure.

\subsection{Position decoding}
\label{subsec:decodingresults}
This section compares the performance of our model with two other models previously used for decoding: the BD model, as explained in \cite{Zhang1998}, and the LP HMM. In BD and LP, positions $X_t$ are used as states (instead of our $S_t$ variables) with state space of size $M$, and in LP (following \cite{Johnson2007}), a transition matrix with rows constrained by Gaussian distributions centered on each position. Maximum likelihood is used for parameter estimation in each.

For these results we used RUN data distinct from that used for parameter estimation (\emph{cross-validation}), and we used the T-maze data since it presents more of a challenge for decoding due to its corners and larger size. We use our fitted model with $\hat{\theta}, \hat{\kappa}$ and the approach to decoding explained in Section \ref{subsubsec:positiondecoding}.

\subsubsection{Decoding comparison: data and performance measures}
\label{subsubsec:decodingmethods}
We used two measures of performance: median decoding error and mean marginal posterior probability. The decoding error of estimate $\hat{x}_t$ we defined as $d \lb x_t, \hat{x}_t \rb$ (the distance function of Section \ref{subsubsec:model}). We then took the median of the decoding errors over all $t$ (rather than the mean since the mean was affected by the heavy tail of the error distribution for all three methods, as shown in Fig. \ref{fig:decodingresults}).

The mean smoothed posterior probability of $x_{1:T}$ is
\begin{equation}
\frac{1}{T}   \sum_{t = 1}^T   p \lb   x_t   \mid   \mathbf{y}_{1:T},   \hat{\theta}   \rb,
\end{equation}
where each term in the sum can be computed with the algorithm in Appendix \ref{app:trajectoryprobability} for our model, or with the forward-backward algorithm for LP. In BD these terms are the single time step posterior probabilities. For an accurate model, the observed trajectory will pass through regions of high posterior probability. Thus, since greater posterior probability on particular positions reduces the posterior variance, a greater value for this measure indicates confidence as well as accuracy, on average, for the decoding method.

\subsubsection{Decoding comparison: results}
\begin{figure*}[t]
\centering
\begin{subfigure}[t]{0.63\textwidth}
\includegraphics[scale=0.22,natwidth=1863,natheight=513]{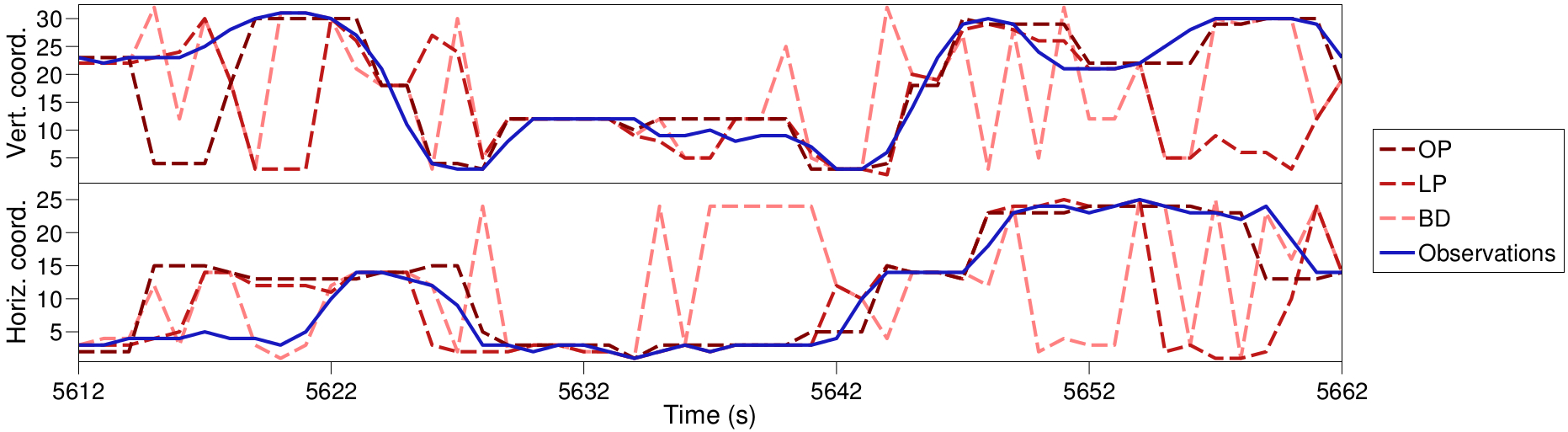}
\end{subfigure}
\hspace{0.5cm}
\begin{subfigure}[t]{0.3\textwidth}
\includegraphics[scale=0.22,natwidth=900,natheight=540]{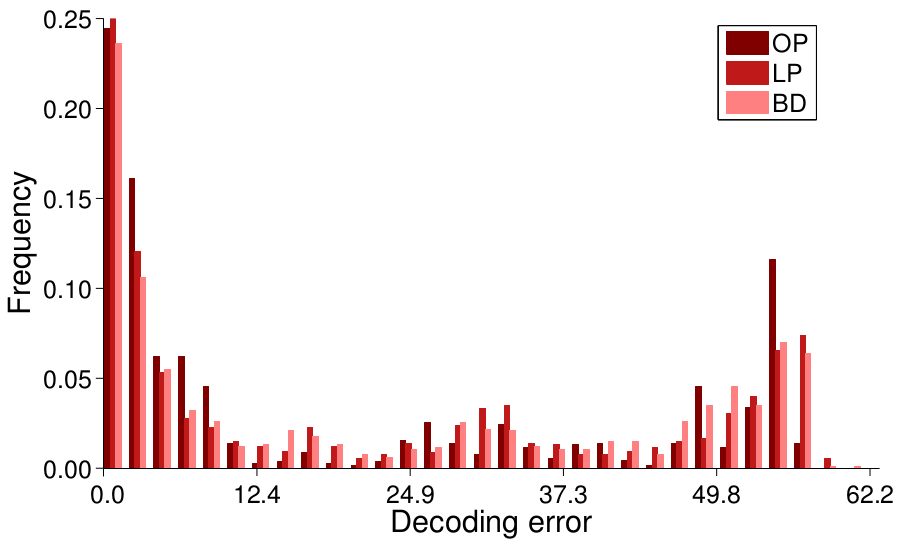}
\end{subfigure}

\begin{subfigure}[t]{0.32\textwidth}
\includegraphics[scale=0.23,natwidth=891,natheight=537]{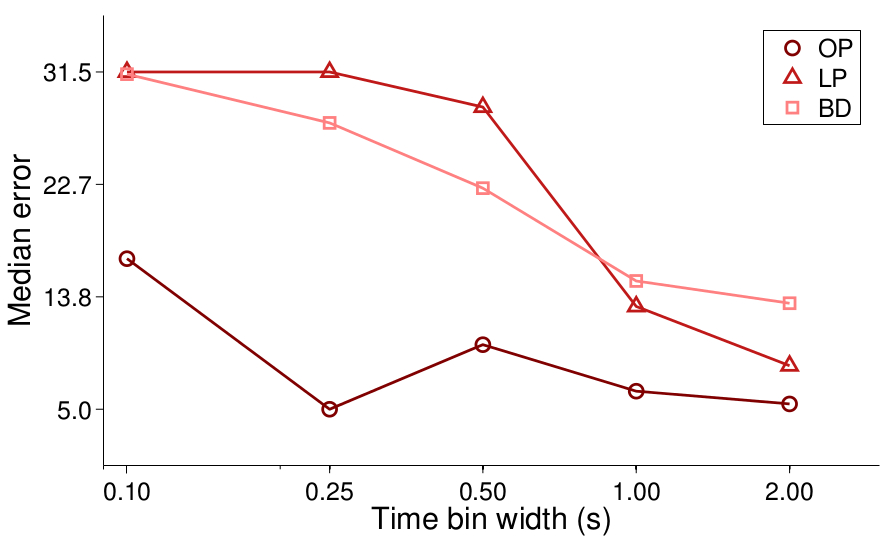}
\end{subfigure}
\begin{subfigure}[t]{0.32\textwidth}
\includegraphics[scale=0.23,natwidth=891,natheight=537]{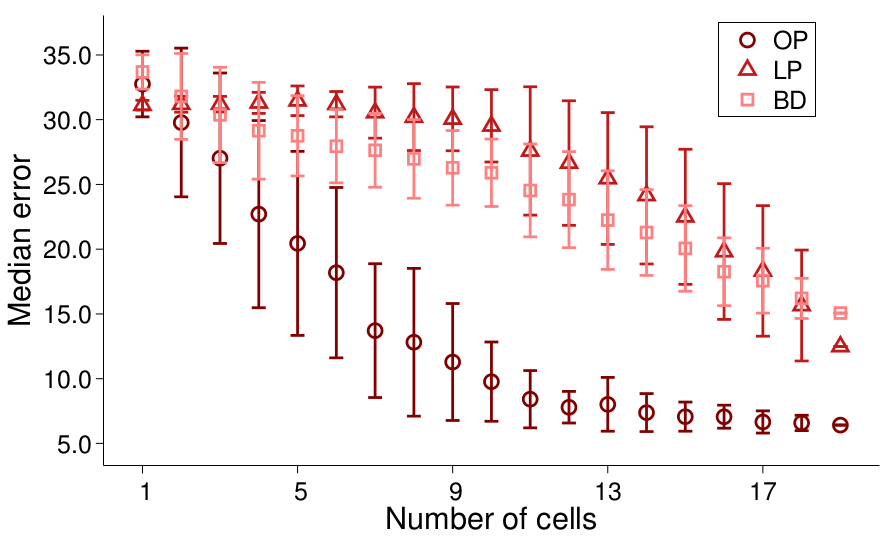}
\end{subfigure}
\begin{subfigure}[t]{0.32\textwidth}
\includegraphics[scale=0.23,natwidth=918,natheight=537]{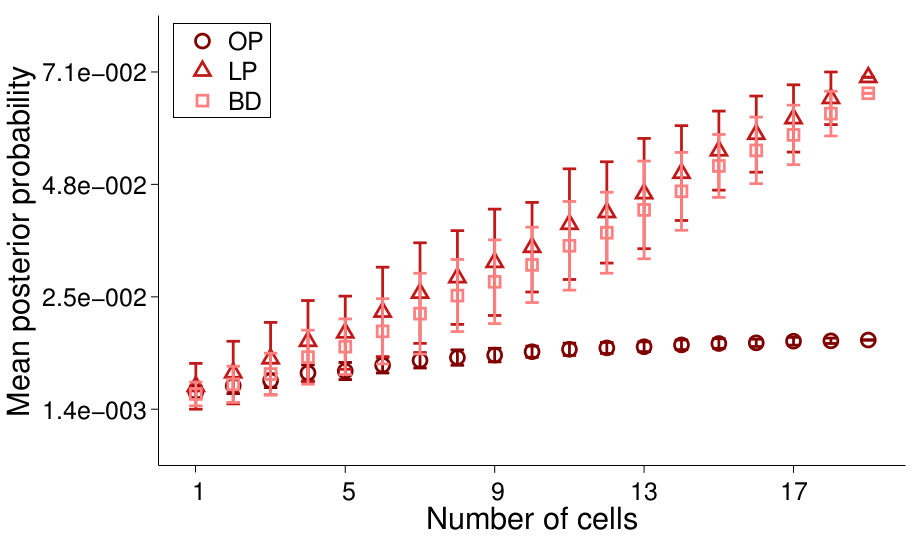}
\end{subfigure}
\caption{Comparison of position decoding performance under our model (OP), the HMM of \cite{Johnson2007} (LP) and the model used in \cite{Zhang1998} (BD). \emph{Top left:} Viterbi estimates of position under each model alongside the observed trajectory (\emph{blue}) in a segment of the T-maze RUN data; $\delta t = 1$s, $C = 19$. \emph{Top right:} empirical distributions of decoding errors (distance between estimated and observed position) for the second half of the T-maze RUN data for the three methods; $\delta t = 1$s, $C = 19$. \emph{Bottom left:} Median decoding error found in same data for a range of values of the temporal bin width $\delta t$. \emph{Bottom centre:} Median decoding error found when subsets of the cell sample were used in decoding. Error bars indicate 1 standard deviation either side of the mean for the subsets used. \emph{Bottom right:} for the same subsets of cells, the mean of the probabilities of the observed position at each time step under the smoothing posteriors computed with each model.}
\label{fig:decodingresults}
\end{figure*}

As per Section \ref{subsubsec:positiondecoding}, we used the Viterbi algorithm to decode position. This is the standard Viterbi algorithm for a HMM for LP, and the algorithm of Appendix \ref{app:viterbiforposition} for OP. For BD the Viterbi estimates are simply the maximum likelihood estimates. A typical interval of the test data is plotted in Fig. \ref{fig:decodingresults}, \emph{top left}, showing each set of decoded estimates alongside observations. The BD estimates have a tendency to jump erratically, whereas the estimates obtained with the HMMs are smoother. Also visible in this figure, towards the end of the interval, is the tendency for the LP estimates to become trapped around one erroneous estimate. This is particularly a problem for small $\delta t$ when it results in massive decoding errors.

For each method we computed the performance measures described in Section \ref{subsubsec:decodingmethods} using models fitted under different values of the parameters $\delta t$ and $C$. For each value of $C$ less than the total number of cells available, $C^\text{max}$, we had a choice of population subset to use; we computed the measures on each subset in a sample of 100 selected at random, or $\binom{C^\text{max}}{C}$ if $\binom{C^\text{max}}{C} < 100$.

These results are presented in Fig. \ref{fig:decodingresults}, \emph{bottom row}. In the bottom left figure is shown how the decoding performance of OP, as quantified by the median error, does not deteriorate drastically with increasing temporal resolution over the range of values of $\delta t$ considered ($2$s, $1$s, $0.5$s, $0.25$s and $0.1$s), unlike LP and BD. The ability of these latter models to decode accurately is severely impaired for $\delta t \le 0.5$s. The median error of decoding and mean posterior probability for varying $C$ are plotted in the bottom centre and bottom right plots, respectively. For these results we fixed $\delta t = 1$s. The error bars in these plots indicate one standard deviation either side of the mean for the cell subsets corresponding to each $C$. We see that in both measures the decoding performance of OP does not degrade much until $C$ is reduced to about $6$ cells, but the performance of LP and BD is badly affected by decreasing $C$. The mean posterior probability of OP is generally lower than for the other models because the posterior variance over positions is generally greater; this is because we do not model positions individually but via a small number of conditional distributions with inherent uncertainty (cf. the position model in Section \ref{subsubsec:model}).

The distribution of decoding errors using estimates obtained with each model, and with $\delta t = 1$s and $C = 19$, is plotted in Fig. \ref{fig:decodingresults}, \emph{top right}. This shows that all three methods suffered from long range errors, but OP did not suffer the very worst errors and had a greater proportion of short range errors than LP and BD. These long range errors are caused by a tendency, in each model, to decode particular positions during times of low firing rates; this is discussed further in Section \ref{subsec:decodingdiscussion}.

\subsection{Replay analysis results}
\label{subsec:replayanalysisresults}
\subsubsection{Simulated data}
\label{subsubsec:replaysimulationresults}
\begin{table*}[t]
\caption{Summary of REST data sets used for replay analysis and results.}
\label{tab:datasetssummary}
\tiny
\begin{tabular*}{1.0\textwidth}{llllllllllll}
\hline
Data set & $\delta t$ (s) & $C$ & $T^{\text{RUN}}$ & $T^{\text{REST}}$ & $\kappa$ & $\hat{\kappa}$ & \#Templates & \parbox{1.3cm}{Mean\\template\\duration (s)} & \#Replay events & \parbox{1cm}{\#SWR\\events} & \parbox{2cm}{Mean SWR\\duration (s)}\\
\hline\\
\parbox{1.4cm}{Sim. linear\\track} & 0.1 & 4 & 10,000 & 10,000 & 4 & 4 & 2 & 48.00 & \parbox{2cm}{39 (of 40,\\$\Omega^* = 20$)} & n/a & n/a \\[0.15cm]
Sim. T-maze & 0.1 & 10 & 10,000 & 10,000 & 5 & 5 & 2 & 25.00 & \parbox{2cm}{30 (of 40,\\$\Omega^* = 20$)} & n/a & n/a \\[0.15cm]
Linear Track & 0.1 & 13 & 9,879 & 9,708 & ? & 5 & 18 & 42.61 & \parbox{2cm}{1,226 ($\Omega^* = 20$)\\316 ($\Omega^* = 150$)} & 261 & 0.07 \\[0.4cm]
T-maze & 0.1 & 19 & 22,569 & 39,943 & ? & 8 & 28 & 25.41 & \parbox{2cm}{8,420 ($\Omega^* = 20$)\\64 ($\Omega^* = 150$)} & 1,492 & 0.07 \\[0.4cm]
\hline
\end{tabular*}
\end{table*}

To assess the performance of our replay detection method on simulated data, we considered replay detection as a binary classification problem where each time bin is to be classified as participating in a replay event or not. First we simulated a REST data set consisting only of spike trains, with 40 known replay events (20 from each of 2 short templates) using the method explained in Section \ref{subsubsec:replaysimulation}. Then, using $\hat{\theta}, \hat{\kappa}$ estimated on the simulated RUN data set (discussed in Section \ref{subsubsec:simulatedparameters}), we applied our replay detection algorithm of Section \ref{subsec:replaydetection} with a range of values for $\Omega^*$, and computed the receiver operating characteristic (ROC) curve parameterised by $\Omega^*$.

Since the ROC curve does not take the rate of false negative classifications into consideration, we also looked at the Jaccard index (\cite{Pang2006}, p74) as an alternative classification measure at each $\Omega^*$ considered. Let $TP$ and $FP$ be respectively the number of true and false positive classifications and let $FN$ be the number of false negative classifications, then the Jaccard index is
\begin{equation}
J \lb \Omega^* \rb   =   \frac{TP}{TP + FP + FN}.
\end{equation}
The maximum value of 1 can only be attained when $FN = 0$, i.e. when no true replay time bins have been misclassified. Thus, the Jaccard index complements the ROC curve by taking into consideration any failure of the algorithm to detect a replay event.

\begin{figure*}[t]
\centering
\begin{subfigure}[t]{0.5\textwidth}
\includegraphics[scale=0.23,natwidth=1386,natheight=518]{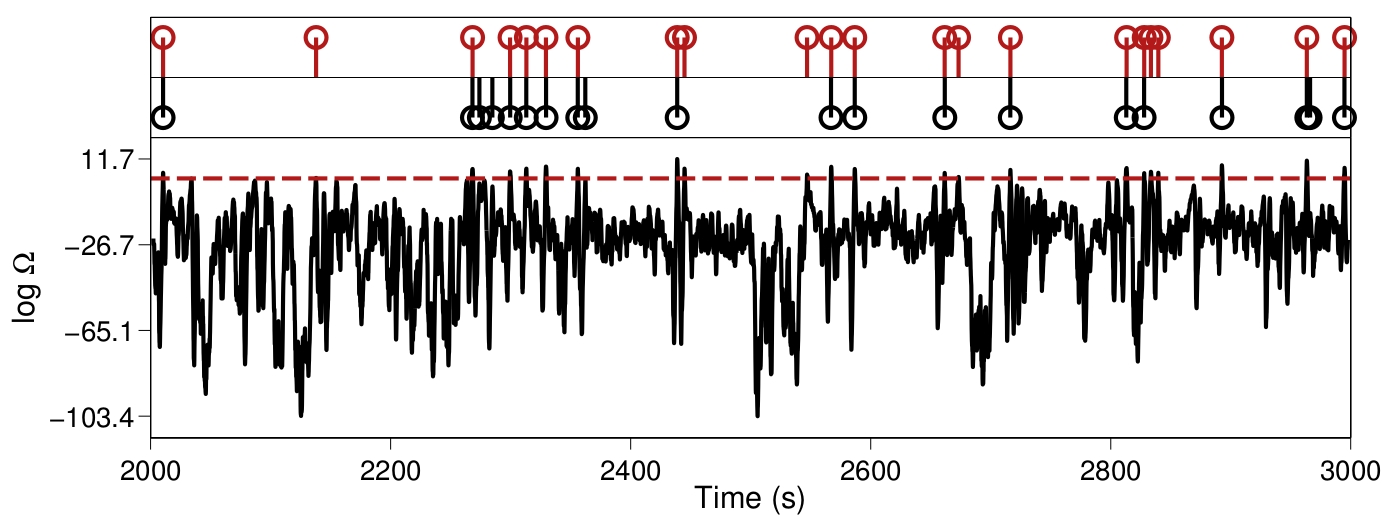}
\end{subfigure}
\begin{subfigure}[t]{0.25\textwidth}
\includegraphics[scale=0.23,natwidth=755,natheight=484]{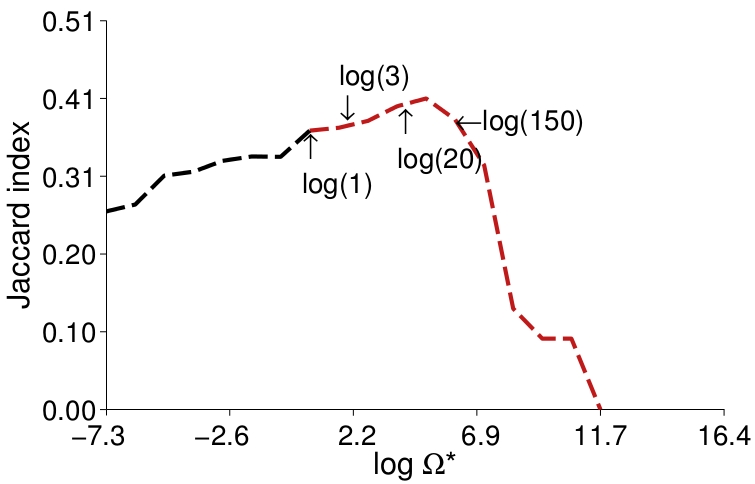}
\end{subfigure}
\begin{subfigure}[t]{0.2\textwidth}
\includegraphics[scale=0.23,natwidth=504,natheight=481]{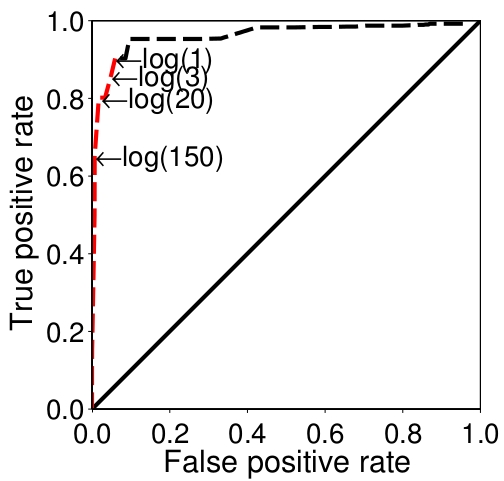}
\end{subfigure}

\begin{subfigure}[t]{0.5\textwidth}
\includegraphics[scale=0.23,natwidth=1370,natheight=518]{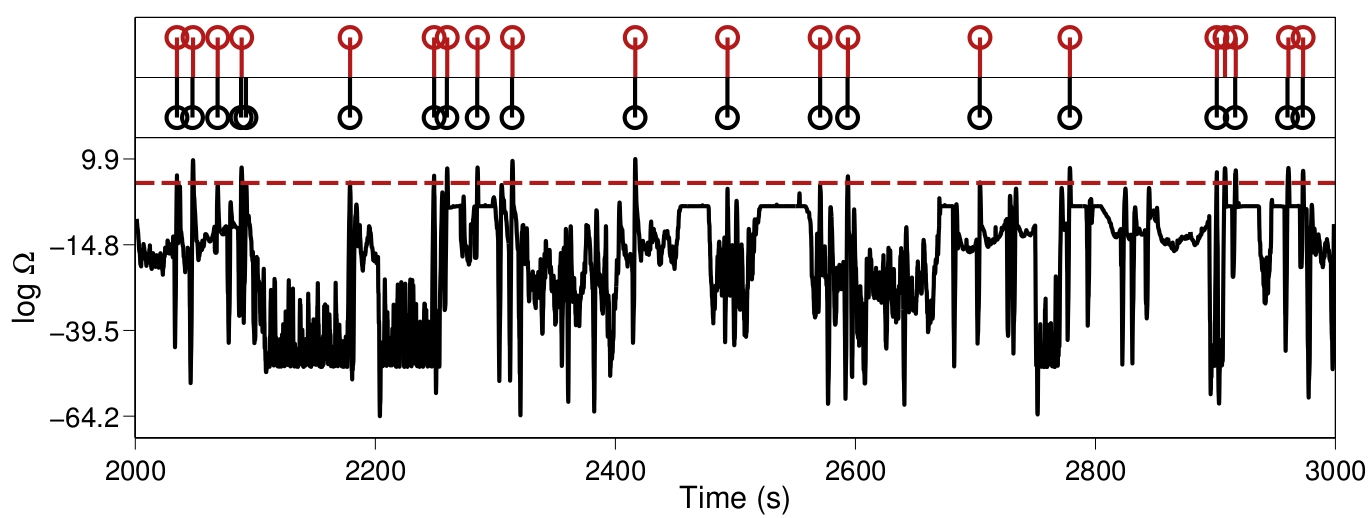}
\end{subfigure}
\begin{subfigure}[t]{0.25\textwidth}
\includegraphics[scale=0.23,natwidth=759,natheight=484]{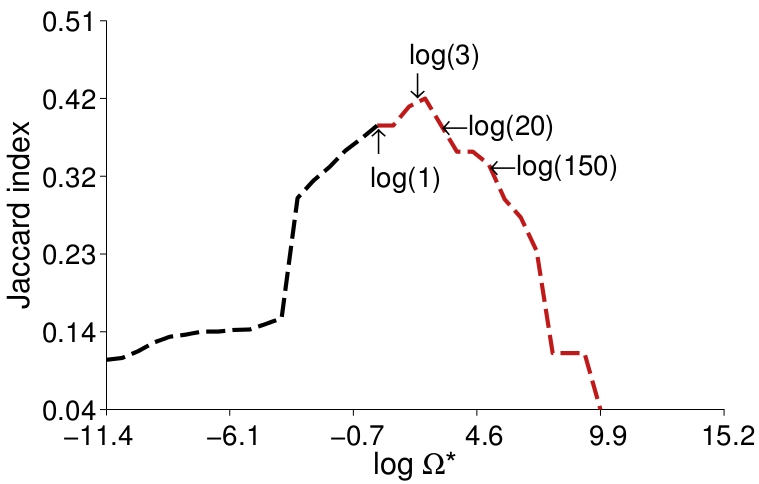}
\end{subfigure}
\begin{subfigure}[t]{0.2\textwidth}
\includegraphics[scale=0.23,natwidth=504,natheight=481]{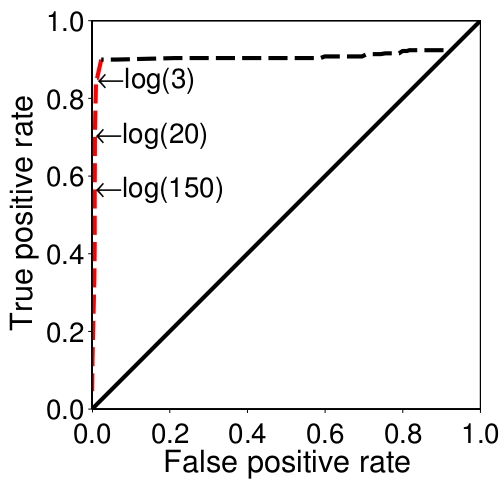}
\end{subfigure}
\caption{Evaluation of replay discovery in simulated data. \emph{Top left, lower panel:} replay score for a template (on the simulated linear track), plotted at the midpoint of the template as it is moved across the data. The red line indicates a threshold of $\Omega^* = 20$. \emph{Top left, upper panel:} red stems indicate times of replay discovery (when a local maximum of the replay score exceeds $\Omega^*$); black stems indicate times of replay events simulated using the method described in Section \ref{subsubsec:replaysimulation}. \emph{Top centre and right:} respectively the Jaccard index curve and ROC curve for discovery of replay of the template considered as binary classification. In each plot the curve is parameterised by $\Omega^*$; the red segment corresponds to $\Omega^* >= 1$. \emph{Bottom row:} similar plots but for a particular template in the simulated T-maze data set.}
\label{fig:simulatedreplayresults}
\end{figure*}

The ROC curve and Jaccard index for the replay detection of one template in each of the simulated data sets are presented in Fig. \ref{fig:simulatedreplayresults}. In both data sets the false positive rate is low ($<5\%$) for $\Omega^* > 1$, with good true positive rates ($>70\%$) for a wide range of $\Omega^*$, and is still about $60\%$ for the conservative $\Omega^* = 150$. The Jaccard index reaches a peak for positive $\Omega^*$ in this range and only starts to decrease beyond $\Omega^* = 150$. Also in Fig. \ref{fig:simulatedreplayresults} are plotted the corresponding profiles of $\Omega$ (as a logarithm, for clarity) and the times of simulated and detected replay for a particular value of $\Omega^*$. It can be seen how the times of replay detection (red stem markers) refer to the times of maxima of $\Omega$ above the threshold. In both data sets most of the replay events are discovered (97.5\% in the linear track, 75\% in the T-maze) with a small number of false positive errors.

\subsubsection{Replay in experimental data}
\label{subsubsec:replayexperimentalresults}
\begin{figure*}[t]
\begin{minipage}[c]{0.6\textwidth}
\begin{subfigure}[t]{0.4\textwidth}
\includegraphics[scale=0.28,natwidth=638,natheight=515]{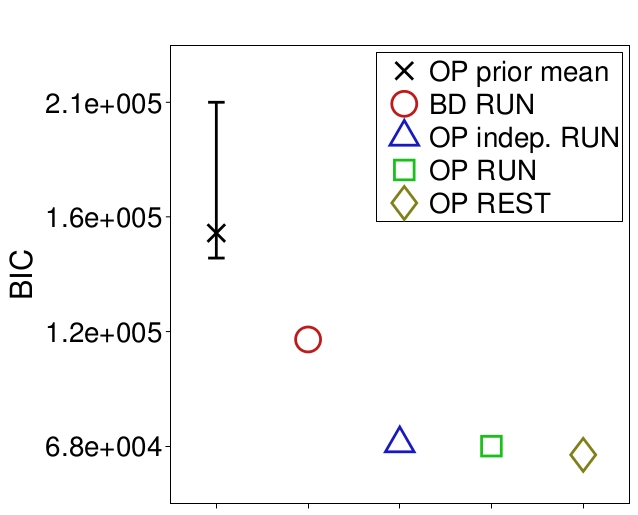}
\end{subfigure}
\hspace{1.0cm}
\begin{subfigure}[t]{0.4\textwidth}
\includegraphics[scale=0.28,natwidth=638,natheight=519]{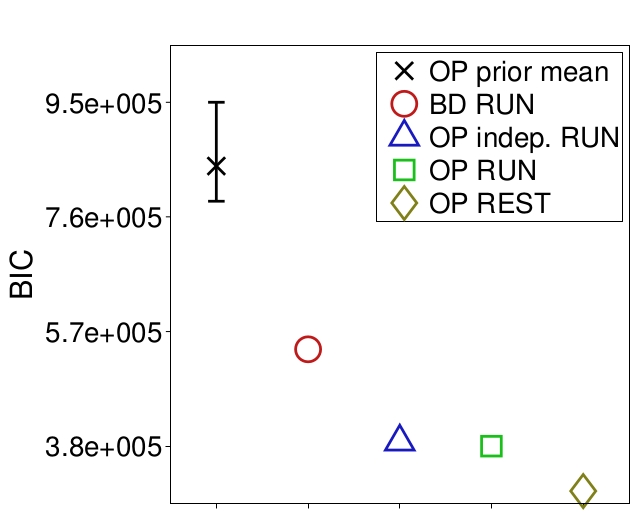}
\end{subfigure}
\end{minipage}\hfill
\begin{minipage}[c]{0.3\textwidth}
\caption{Bayesian information criteria (BIC) for model fit assessment on the REST data, used for replay analysis. The green square represents our model (OP) with parameter values fitted to RUN data. This we compare against: parameterisations of the OP sampled from the prior for $\theta$ (error bars indicate the 5th and 95th percentiles of the sample) and the expectation of the BIC over the prior, the Bayesian decoder fitted to RUN, the OP fitted to RUN but with its Markov chain dynamics replaced with a time invariant distribution over states, and the OP fitted directly to the REST data. \emph{Left:} linear track data, \emph{right:} T-maze data.}
\label{fig:modelcomparisonresults}
\end{minipage}
\end{figure*}

We applied the algorithm of Section \ref{subsec:replaydetection} to our experimental REST data sets using $\hat{\theta}$ estimated from RUN data. First we used the model comparison approach described in Section \ref{subsubsec:modelcomparison} to verify that the model with parameter values $\hat{\mathbf{\theta}}$ was a good fit to the REST data in both data sets. As shown in Fig. \ref{fig:modelcomparisonresults}, the BIC on the REST data for our model with $\hat{\theta}$ estimated from RUN data (OP, RUN, \emph{green square}) is close to the benchmark parameterisation - the model fitted to the REST data directly (OP, REST, \emph{gold diamond}) - relative to the model with $\theta$ sampled from the prior and the prior mean (\emph{black cross}). We draw reassurance from this that the model with parameterisation $\hat{\theta}$ learned from RUN is a good fit to the REST data used for the replay analysis.

This is further supported by OP (RUN) having a lower BIC than the similar model parameterisation with independent rather than Markovian dynamics (Section \ref{subsubsec:modelcomparison}), also shown in Fig. \ref{fig:modelcomparisonresults}. Thus, the dynamics from RUN, as characterised by $\hat{\mathbf{P}}$, persist in REST and are described well by $\hat{\mathbf{P}}$. We also compare the BIC on REST data of OP (RUN) with that of BD, fit to RUN. We find that the former is much lower, both with and without Markovian dynamics, implying that with its smaller state space, our model is a more parsimonious characterisation of the data.

\begin{figure*}[t]
\centering
\includegraphics[scale=0.38,natwidth=1641,natheight=409]{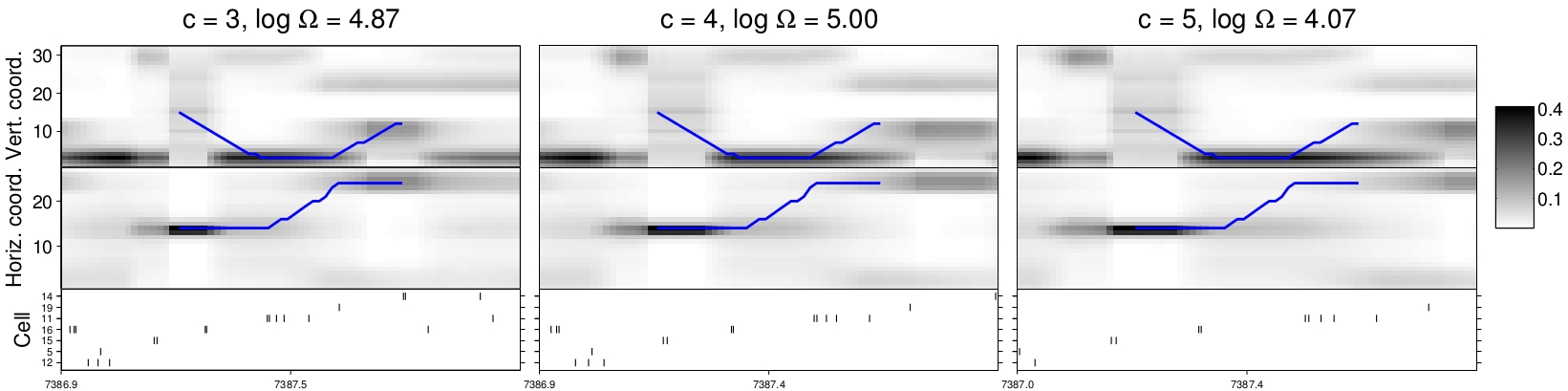}\\
\vspace{1em}
\begin{minipage}[c]{0.45\textwidth}
\centering
\includegraphics[scale=0.38,natwidth=640,natheight=372]{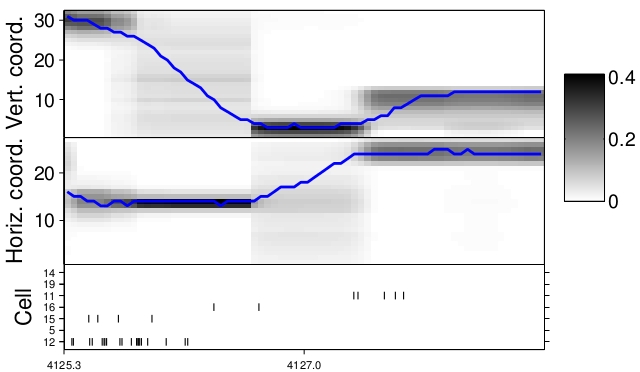}
\end{minipage}
\begin{minipage}[c]{0.45\textwidth}
\caption{Example of a replay event discovered in experimental data. Each subfigure depicts a time interval around a discovered replay event. In the top two panels are plotted, at each time step, the smoothing posteriors over position (obtained using Eq. \eqref{eq:positionposterior}), marginalised to the vertical and horizontal position coordinates, with a greyscale shade indicating probability. A blue line indicates the template trajectory. The bottom panels depict a subset of the concurrent spike trains as a raster of spike times: only cells that spiked during the interval are represented. \emph{Top row of subfigures:} the same replay event as discovered in the T-maze REST data at compression rates 3, 4 and 5; the peak replay score was observed for this event at a compression rate of 4. \emph{Left:} a similar interval around a discovery of the same template in the T-maze RUN data. Here the blue line describes the observed trajectory.}
\label{fig:replayexample}
\end{minipage}
\end{figure*}

In order to demonstrate more explicitly how our replay detection works, in Fig. \ref{fig:replayexample} is depicted an example of a detected replay event of a template in the T-maze data. This template comprises a path around the forced turn and into a rest area. The images depict the smoothed posterior distributions over position (with greyscale shade indicating probability mass), marginally for the two spatial dimensions, at each time step in an interval around the event. The top row of the figure shows the replay event at three consecutive compression rates $c$, with the central panel showing the event detected with peak $\Omega$ at compression rate $c = 4$. Also plotted is the template trajectory, in blue, and a raster of spike times for all cells that spiked during the interval. Regions of high posterior probability follow the template, and greater $\Omega$ corresponds to a closer fit of the template to the position posteriors. Below and to the left of the figure is plotted an example of the same template being matched against an interval of uncompressed RUN data, now with the observed trajectory depicted in blue. We see a similar trajectory of peak posterior probability tracking the observed trajectory, which gives us confirmation (by eye) that the episode detected in REST matches an encoded RUN experience. We also see in this interval of RUN a similar pattern of spike trains from the same cells as in the replay event.

\begin{figure}[t]
\begin{minipage}[c]{0.5\textwidth}
\centering
\begin{subfigure}[h]{0.48\textwidth}
\includegraphics[scale=0.16,natwidth=950,natheight=934]{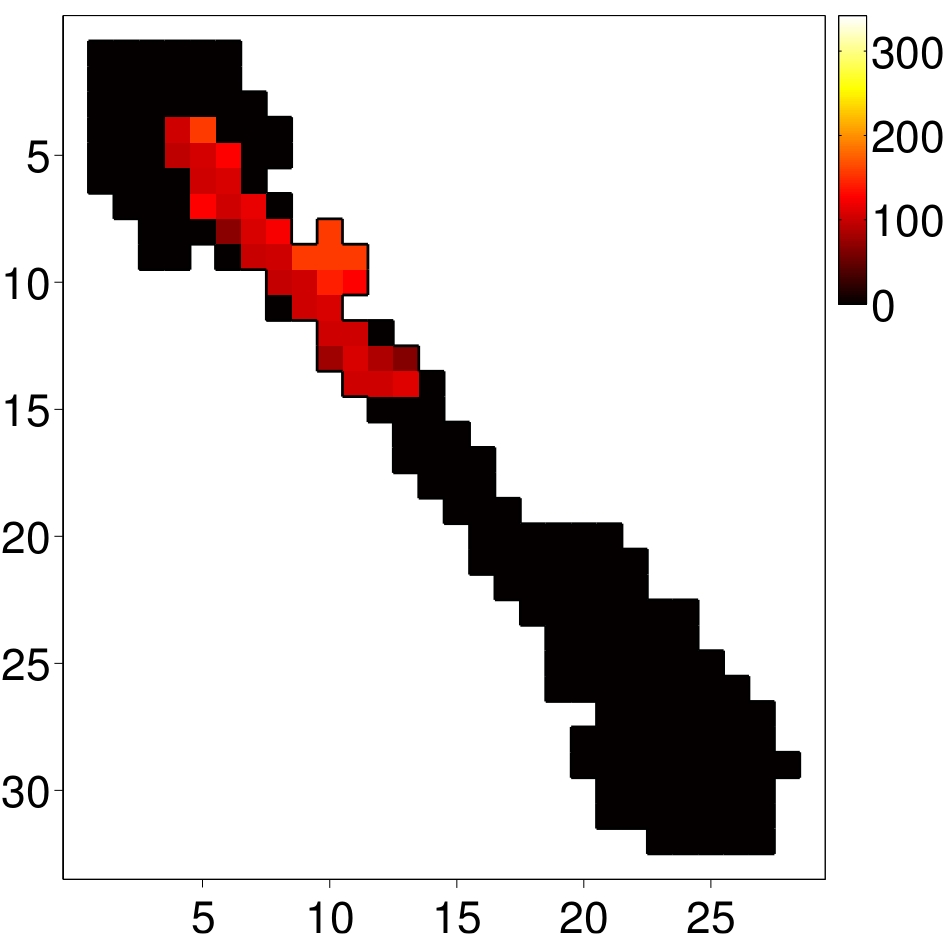}
\label{subfig:kenanoutbound}
\end{subfigure}
\begin{subfigure}[h]{0.48\textwidth}
\includegraphics[scale=0.16,natwidth=950,natheight=934]{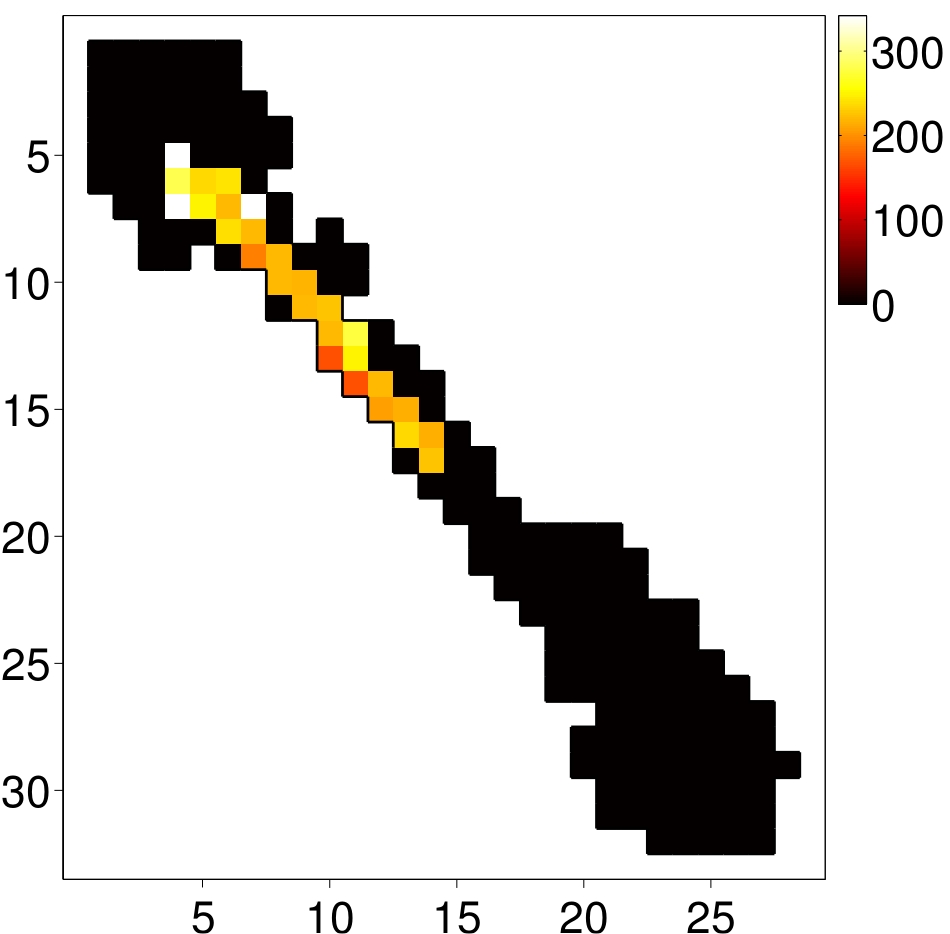}
\label{subfig:kenaninbound}
\end{subfigure}

\begin{subfigure}[h]{0.48\textwidth}
\includegraphics[scale=0.16,natwidth=925,natheight=934]{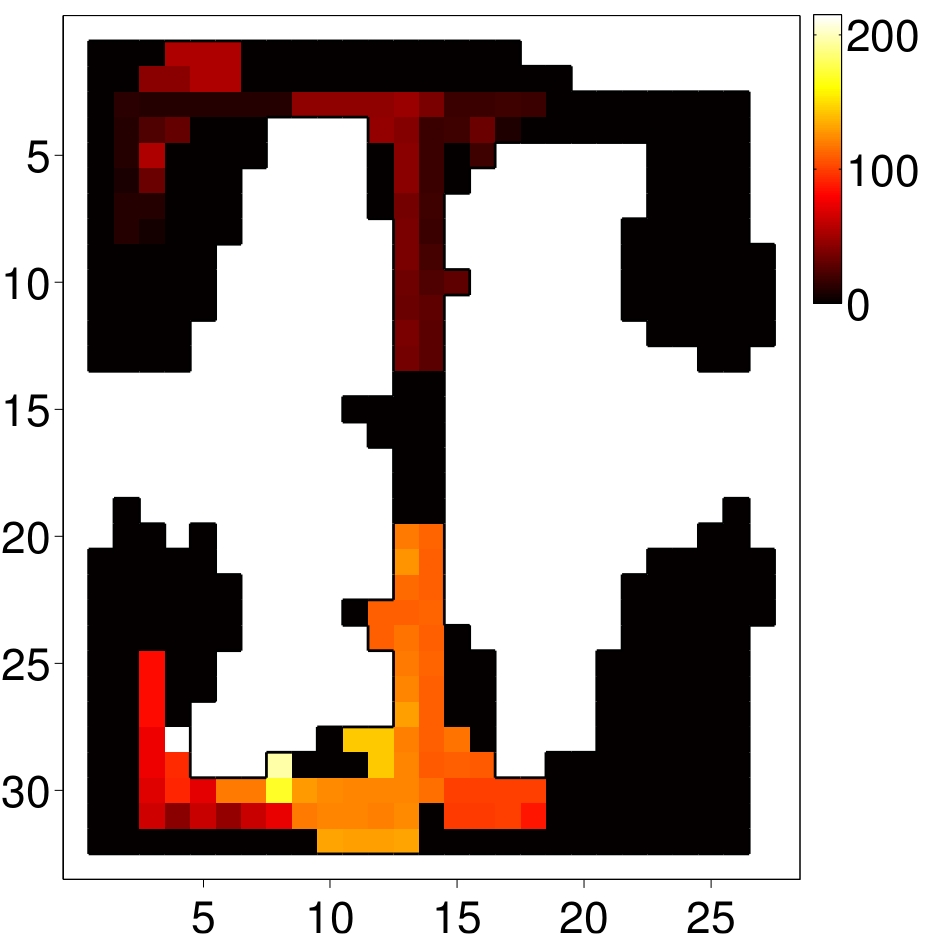}
\label{subfig:jasperoutbound}
\end{subfigure}
\begin{subfigure}[h]{0.48\textwidth}
\includegraphics[scale=0.16,natwidth=925,natheight=934]{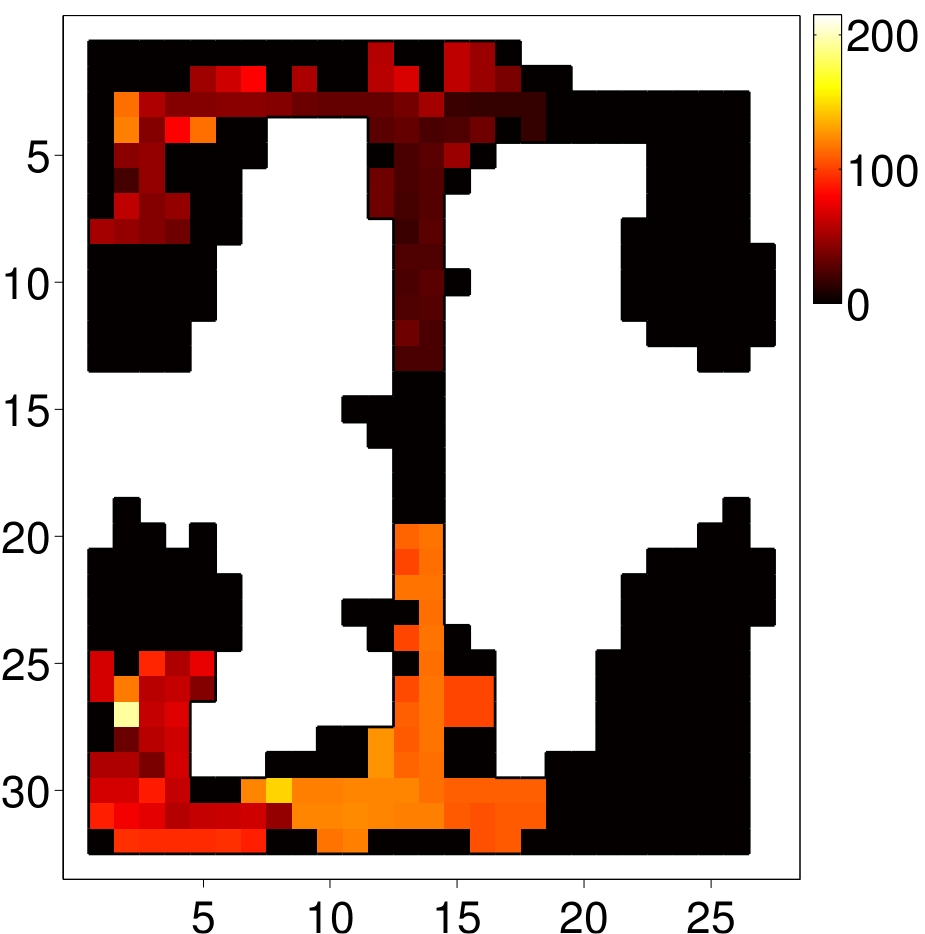}
\label{subfig:jasperinbound}
\end{subfigure}
\end{minipage}\hfill
\begin{minipage}[c]{0.4\textwidth}
\centering
\caption{Normalised frequency of replay ``visits'' to each position; replay events found using a threshold of $\Omega^* = 20$ in the algorithm of \ref{subsec:replaydetection}. Colours indicate the number of times a discrete position formed part of a template that was replayed, normalised by the number of templates that include that position. Results have been split between outbound templates, heading away from the centre of the environment, and inbound templates, heading towards the centre. \emph{Top left:} linear track, outbound templates. \emph{Top right:} linear track, inbound templates. \emph{Bottom left:} T-maze, outbound. \emph{Bottom right:} T-maze, inbound. The choice end of the T-maze is at the bottom.}
\label{fig:replaylocationresults}
\end{minipage}
\end{figure}

Details of our replay analysis are summarised in Table \ref{tab:datasetssummary}. Using a conservative threshold of $\Omega^* = 150$, we found 316 and 64 events in the linear track and T-maze data sets respectively. These numbers are on the same order as those reported in other replay studies using a range of methods; for example, \cite{Ji2006} found about 39 candidate events (not restricted to those in SWRs) per session, \cite{Lee2002a} found 57 events (based on triplet sequences of cell activation) between three rats, and \cite{Nadasdy1999} found up to 40 events (repeats of spike sequences) per session.

Fig. \ref{fig:replaylocationresults} depicts the location in each environment of detected replay events as rates of participation of discrete positions in replayed template trajectories. In the linear track, more than twice as many replay events were of templates that started at the top end of the track (from the perspective of Fig. \ref{fig:replaylocationresults}) and ended near the centre than in reverse. No replay events were found for the opposite end of the track in either direction. In the T-maze, the most often replayed region was around the ``T'' junction at which only a correct choice resulted in reward. There was also a small preference for right over left turns. No replay events were found for routes into the rest sites in the right half of the environment. There were no easily discernible preferences to replay trajectories either away from or towards the central corridor.

\subsection{Correlation of replay events with hippocampal SWRs}
\label{subsubsec:replayripplecorrelationresults}
\begin{figure*}[t]
\centering
\begin{minipage}[c]{0.6\textwidth}
\begin{subfigure}[h]{0.23\textwidth}
\includegraphics[scale=0.23,natwidth=577,natheight=523]{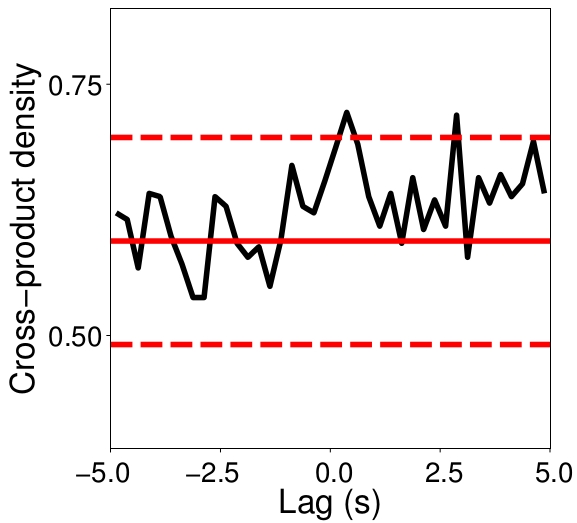}
\label{subfig:kenancrossprodall}
\end{subfigure}
\hfill
\begin{subfigure}[h]{0.23\textwidth}
\includegraphics[scale=0.23,natwidth=577,natheight=523]{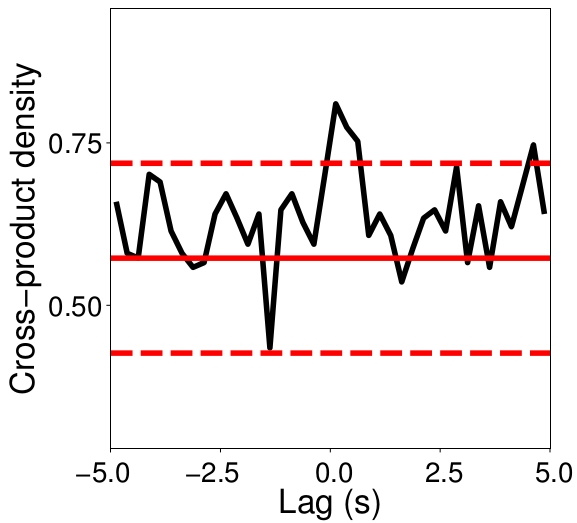}
\label{subfig:kenancrossprodhalf1}
\end{subfigure}
\hfill
\begin{subfigure}[h]{0.23\textwidth}
\includegraphics[scale=0.23,natwidth=577,natheight=523]{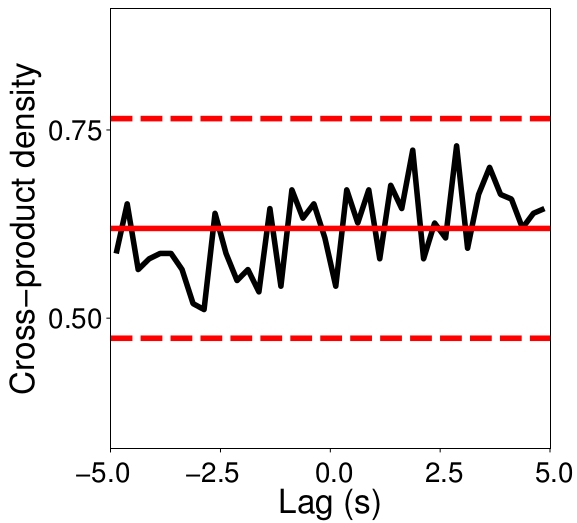}
\label{subfig:kenancrossprodhalf2}
\end{subfigure}

\begin{subfigure}[h]{0.23\textwidth}
\includegraphics[scale=0.23,natwidth=562,natheight=523]{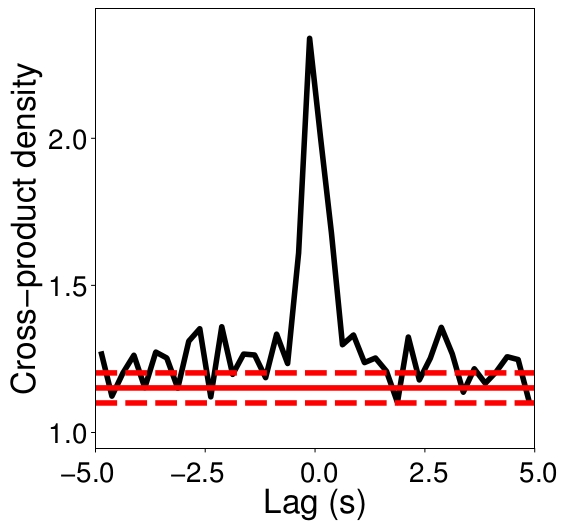}
\label{subfig:jaspercrossprodall}
\end{subfigure}
\hfill
\begin{subfigure}[h]{0.23\textwidth}
\includegraphics[scale=0.23,natwidth=562,natheight=523]{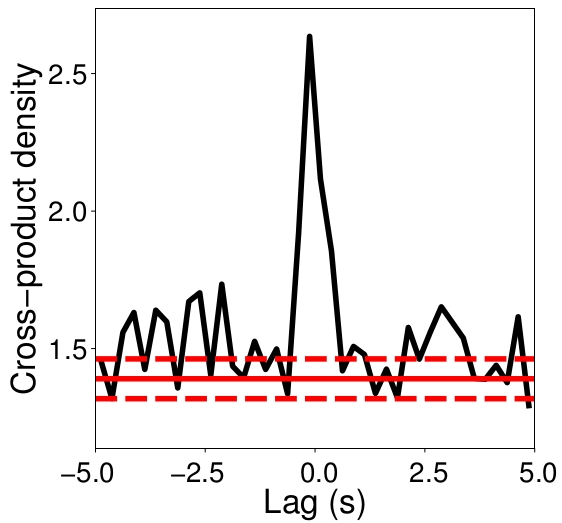}
\label{subfig:jaspercrossprodhalf1}
\end{subfigure}
\hfill
\begin{subfigure}[h]{0.23\textwidth}
\includegraphics[scale=0.23,natwidth=562,natheight=523]{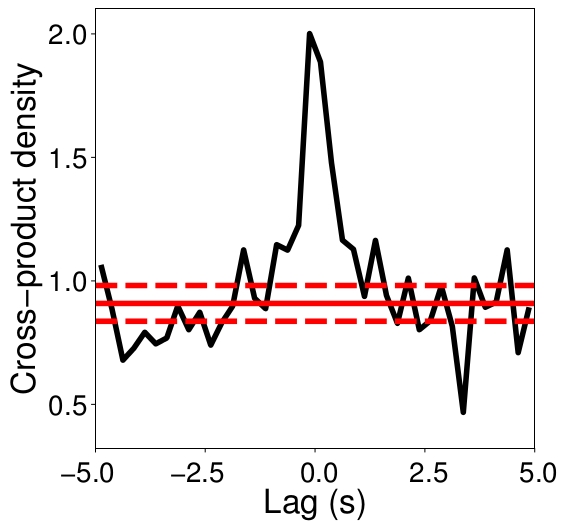}
\label{subfig:jaspercrossprodhalf2}
\end{subfigure}
\end{minipage}\hfill
\begin{minipage}[c]{0.3\textwidth}
\caption{Estimates of the cross-product density $\sqrt(\rho_{rep, rip} \lb   u   \rb)$ from times of SWR events to times of detected replay events at lags $u$ around $0$, obtained from the cross correlogram, $J_{rep, rip} \lb   u   \rb$. Estimates have been square root transformed for variance stabilisation, as in \cite{Brillinger1976}. Solid red lines indicate $\sqrt(\rho_{rep} \rho_{rip})$, the value expected for two independent processes. Dashed red lines indicate approximate confidence limits constructed using a significance level of $\alpha = 0.178\%$, which includes the Bonferroni correction for comparing the estimate at each lag. \emph{Top row:} linear track, \emph{bottom row:} T-maze. \emph{Left:} all REST data used, \emph{middle:} first half of data, \emph{right:} second half.}
\label{fig:crossproductdensityplots}
\end{minipage}
\end{figure*}

We used the methods described in Section \ref{subsec:replayripplecorrelation} to identify SWR events in the LFP recorded during REST for each data set (summarised in Table \ref{tab:datasetssummary}). We computed the cross correlogram for the times of SWR events and replay events, using a bin width of $\tau = 0.25$s, appropriate to the $\delta t$ used and the average duration of SWR events. As explained in Section \ref{subsec:replayripplecorrelation}, this is an unbiased estimator of the second-order product density function, $\rho_{rep, rip} \lb   u   \rb$. Values of $\sqrt{   \hat{\rho}_{rep, rip} \lb   u   \rb   }$ are plotted in Fig. \ref{fig:crossproductdensityplots}, between $-5$s and $5$s. An approximate $0.178\%$ confidence interval, which includes a Bonferroni correction for multiple comparisons, is plotted around the value for $\rho_{rep, rip} \lb   u   \rb$ under the assumption of no correlation, to highlight deviations from it as peaks or troughs outside of the interval. The interval is wider for the linear track results because there fewer events were detected (likely due to shorter recordings, i.e. smaller $T$).

We observe a significant peak around zero for both the linear track and T-maze data sets (Fig. \ref{fig:crossproductdensityplots}, \emph{left column}), from which we conclude that the times of replay events and SWR events coincide. The peak around zero extends into positive lags more than negative lags, signifying that the SWR events occur first (cf. Eq. \eqref{eq:crosscorrelogram}) as would be expected if replay occurs during ripples. Regarding peaks away from zero we must consider that estimates of $\rho_{rep, rip} \lb   u   \rb$ become less reliable as the lag $|u|$ increases (\cite{Brillinger1976}). The results presented in Fig. \ref{fig:crossproductdensityplots} were based on the replay events detected using a threshold of $\Omega^* = 20$. Using the more conservative threshold of $\Omega^* = 150$ we draw the same conclusions, except in the case of the linear track data for which we did not have enough events to demonstrate a significant correlation.

We defined the second-order product density function for \emph{stationary} processes. In order to guard against deviations from stationarity affecting our results, we performed the same analyses on events detected in subsections of REST. These are plotted in Fig. \ref{fig:crossproductdensityplots}, \emph{middle} and \emph{right}. We find that the correlation between the processes persists at this finer scale in the T-maze data. No significant correlation is found in the second half of the linear track data, but the correlation does exist in the first half, so the correlation does persist across different scales in at least part of the data.

%% file: discussion.tex
\section{Discussion}
\label{sec:discussion}
\subsection{Improvements afforded by our model}
In developing our model, we recognised the advantages of the statistical modelling approach to spike train analysis: that sources of variation in observation variables are explicitly accounted for, enabling one to quantify the probability of outcomes and make predictions. Furthermore, we recognised the advantage of including dynamics via the HMM framework, as undertaken by \cite{Brown1998} and used for replay analysis in \cite{Johnson2007}, for the accurate characterisation of data with clear dependence through time.

By removing position observations from the hidden process - the approach of LP - out to an observed process parallel to the spike trains (cf. Fig. \ref{fig:dag}), we achieve two important improvements. Firstly, we elicit from the data itself structure around the trajectory of the animal and how this relates to the spike trains, within the constraints imposed by our model distributions. This structure is described by the number, location and shape of broad regions of the environment that are, to the extent permitted by the data, the smallest regions discernable by variation in the spike trains. We bring to this inference no prior knowledge, using uninformative priors as far as possible, including our inference for the number of states, thus allowing the data to ``speak for itself''.

Secondly, the disassociation of discrete positions from states of the model, which reduces the number of parameters to the small set necessary for our coarse-grained representation of space. This parsimony is confirmed by the lower BIC for OP than BD (cf. Fig. \ref{fig:modelcomparisonresults}). This makes it easier to make robust estimates of the parameters with limited data, and, by performing inference for the number of states and for the position model parameters, we are able to explore the neuronal ensemble's representation of space via the number, size and shape of these regions; this is demonstrated in Fig. \ref{fig:fittedpositiondistributions}. Further opportunities are provided for studying the brain's representation of space by performing these inferences under different experimental conditions, such as different stages of the animal's training or familiarity with the environment.

\subsection{Decoding performance of the model}
\label{subsec:decodingdiscussion}
There are two important consequences for our model of the decoding analysis presented in Section \ref{subsec:decodingresults}. Firstly that the catastrophic rate of decoding error we observe with BD and LP at high temporal resolution does not occur with OP. This means we are able to use a greater resolution at the parameter estimation stage and thereby capture variations in the spike count that occur on a more precise time scale.

The reason for this benefit seems to be OP's coarse-graining of position to a small number of minimally discernable regions. The problem seen in BD and LP has to do with an unwanted feature of these kinds of model: that it implies some positions are encoded by the absence of spikes. \cite{Zhang1998} noted decoding errors in the form of large jumps or discontinuities in decoded trajectories, mostly occurring when the animal was still and firing rates were low. It appears from their Fig. 3 that these erroneous decoded estimates were of a small number of particular positions. This has also been our experience using these methods, in particular at high time resolution, as exhibited by the jumps in decoded trajectory in the top left plot of Fig. \ref{fig:decodingresults}. We have also observed trajectories decoded using LP getting trapped in particular locations at high time resolution ($\Delta < 1$s). In both BD and LP, particular positions maximise the likelihood (conditional probability of spike train observations given position) for low spike counts, and so will maximise, or at least strengthen, the posterior distribution over these positions, and hence they will be decoded with methods based on the likelihood. 

In OP, however, a particular state will maximise the likelihood for low spike count observations, but these periods are brief relative to the jump rate of the Markov chain due to the relatively small number of states, and so these observations will not have such an overpowering effect on the posterior. Thus, the consequence of positions encoding inactivity are avoided in OP by its association of broad regions, rather than discrete positions, with states of the model, and by eliciting the details of these regions from the data itself.

The second advantage conferred by OP as demonstrated by our decoding results is that we can achieve good results with a small number of cells (little degradation in decoding performance for a sample of $6$ or $7$ cells compared with $19$ cells). This makes our model a good choice for decoding with limited data, as may be the case when we wish to record from a particularly idiosyncratic or sparse population of neurons, when recordings are of poor quality and cannot be easily clustered, or when less advanced equipment is available.

More than simply as a tool for inferring the information content encoded in spike trains, decoding using the posterior distribution (including Viterbi estimates) can be seen as a posterior predictive validation of the model (\cite{Gelman2003}, p188); that is, as a means for veryifying the statistical model's characterisation of the encoding of position in spike trains. The decoding results thus support our model as being useful for the study of replay, and, since our method for replay detection is based on the same principle as the decoding algorithm - that of using the posterior distribution over position to infer the information content of spike trains - the advantages demonstrated for our model in decoding also apply to replay detection. 

\subsection{The SMC algorithm}
\label{subsec:smcdiscussion}
Other solutions to the identifiability problem in HMMs have been proposed, but these come with their own issues. As discussed in \cite{Scott2002}, these often involve imposing structure on the prior distribution of exchangeable parameters, or otherwise breaking the symmetry in the model. This kind of solution is difficult to justify when there is no a priori reason to bias parameters away from each other or impose constraints on, for example, the ordering of parameters such as mean firing rates, and inferences may be influenced by the choice of constraint. The solution presented in \cite{Chopin2007} and used here does not require any such constraints, permits a fully Bayesian approach to parameter inference with uninformative priors and facilitates a Bayesian approach to model selection that accomplishes the task of eliciting from the data itself the required complexity for the spatial representation. Inference for parameters is subject to sampling error, but targets the true values, unlike in the methods of \cite{Chen2012}, and we can achieve an arbitrary degree of accuracy by increasing the particle sample size, constrained only by computer resources.

\subsection{Use of the BIC for model comparison on REST data}
We chose to use a likelihood-based technique for model comparison, the BIC, to verify that the model fitted to RUN data was a good fit for the REST data. It was important for our application of replay detection using an evaluation of the posterior as in Eq. \eqref{eq:replayscore} that we assess the fit of a particular parameterisation of the model - the posterior mean estimate $\hat{\theta}$, in particular - rather than the model fit marginal of model parameters as is typically done in a Bayesian model comparison, for example with the deviance information criterion (DIC, cf. \cite{Gelman2003}, p183) and the Bayes factor (\cite{Kass1995}). Furthermore, our task was not merely to demonstrate the general out-of-sample predictive power of our model, as is achieved with the DIC, but predictive power specifically on the REST data. The BIC is useful for this because it can be computed using the REST data likelihood. The BIC also permits comparisons between non-nested models, for example between OP and BD, and its inclusion of a penalty for model complexity provides a stronger test for OP against the time-independent alternative (which has no transition matrix).

\subsection{Replay detection methods}
The motivation for our approach to replay detection was to take further the model-based, decoding approach used profitably in other studies, and by so doing overcome the principle challenges associated with replay detection and enable a more extensive analysis of the phenomenon. Whereas studies such as \cite{Johnson2007} have used the time marginal posterior distributions of position given spikes, discussed in Section \ref{subsubsec:positiondecoding}, we use the posterior distribution over \emph{trajectories}: sequences of position random variables considered jointly (cf. Eq. \eqref{eq:replayscore}). The neuronal representations we wish to identify in replay detection are dynamic: their temporal dependence structure is essential. It is therefore important for the detection of replay with a model of the relevant processes that one starts from the most general characterisation permitted, so one does not make any inappropriate assumptions (of independence, for example) that make the model itself appear inadequate. Indeed, we saw by comparison of the BIC in Fig. \ref{fig:modelcomparisonresults} that the model with Markovian dependence was a better fit to spike train data than the same model with temporal independence.

Furthermore, in studies of replay such as \cite{Davidson2009}, a model is used to decode a trajectory in the sense of computing a point estimate, which is then tested against criteria that constitute an operational definition of replay. Our advancement is to recognise in the model a description of all trajectories that \emph{might} be encoded (the posterior over positions given spike trains, i.e. the numerator of Eq. \eqref{eq:replayscore}). We thus make full use of the information contained in the posterior distribution rather than only taking from it a point estimate.

As well as specifying criteria for replay detection, other authors have found it necessary to guard against mistakenly detecting replay by chance (a type I error in the language of hypothesis testing). To this end, \cite{Davidson2009}, \cite{Dragoi2011} and \cite{Pfeiffer2013} used informal hypothesis testing to demonstrate positive discovery at a nominated statistical significance level. These tests are informal since the distribution of their test statistic (typically a replay score methodologically equivalent to our $\Omega$) under the null hypothesis (of no replay) is unknown, and hence it is not clear how to calculate a $p$-value. This is resolved in these studies by the use of a permutation test (or ``Exact test'', \cite{Good2005}), in which the unknown distribution is arrived at simply by evaluating the test statistic under all possible permutations of the test data. Since this is infeasible for candidate replay events of nontrivial length, a Monte Carlo version is typically used, in which a random sample of the test statistic is obtained via shuffling procedures on the test data. This approach comes with its own uncertainty: the ``Monte Carlo $p$-value'' is an approximate $p$-value when the sample taken is not exhaustive.

In our method, the risk of mistaking chance observations for true replay is accounted for by the marginal distribution over trajectories, e.g. $p \lb   X_t = x_1,   \dots,   X_{t + a - 1} = x_a   \mid   \mathbf{\theta}   \rb$ for a template of length $a$ at offset $t$; cf. Eq. \eqref{eq:replayscore}. Setting a positive threshold for $\Omega$ protects against trajectories that may be probable a posteriori due to a bias in the model favouring those trajectories; we must have $\Omega > \Omega^*$ only when a trajectory is decoded above ``chance'' as represented by the marginal distribution. We do not need to resort to ad-hoc tests of statistical significance or the kind of shuffling procedures mentioned above, which have an element of subjective judgement in their design, nor do we need to accept any approximate $p$-values of uncertain accuracy.

\subsection{Limitations of the template matching approach}
For the results presented in Section \ref{subsubsec:replayexperimentalresults} we used segments of an observed trajectory (i.e. from RUN data) as templates for replay detection. However, the decision of which segments to use was arbitrary, and was guided only by our interest in particular regions of the environment. This means we are unable to determine the true start and end times of a replay event, which also precludes us from drawing conclusions regarding the relationship between replay event duration and time compression rate.

It may be possible to combine our replay analysis methods with the decoding algorithm to make a more comprehensive study of what is being replayed and at what compression that does not depend on our choice of templates, for instance by eliciting replayed trajectories directly from the data such as segments of the Viterbi path during REST.

\section{Conclusion}
We have presented a dynamic statistical model relating multiple parallel spike trains to concurrent position observations that explains the data in terms of discrete levels of spiking activity and broad regions of an environment, corresponding to distinct states of a Markov chain. We have seen an improvement in decoding performance over other models which seem to be consequences of our use of states distinct from individual positions. In this way our model improves upon those of \cite{Brown1998} and \cite{Zhang1998}, used in most recent studies of replay, in which positions are identified with states of the model. The approach taken to model fitting achieves Bayesian inference for parameters, overcoming the model identifiability problem suffered by HMMs with a likelihood invariant to permutations of the state, while also performing Bayesian inference for model size.

We have also presented a new model-based method for the analysis of replay in spike trains, and demonstrated how this can be employed with our model to discover replayed representations of position trajectories of arbitrary length and content. We have argued that consideration of the model likelihood, and how it compares with certain benchmarks, is an appropriate way to demonstrate a model as being an appropriate characterisation of data distinct from that used for parameter inference. Once this is established, our method for identifying replay is to compare the posterior probability of a specified trajectory segment given the spike trains intended for analysis with the marginal probability of the trajectory segment, and identify times at which the posterior probability obtains large maxima. Post hoc tests of significance are not required since variability in trajectories is captured by our model.

The methods presented here are well-suited to the study of replay even in problematic data conditions such as small neuronal sample size. With further scope for development, in particular in respect to the way we construct template trajectories for detection, we propose to use these methods to explore the open questions about the phenomenon of replay, such as the role of time compression, the details of replay episodes of varying temporal and spatial characteristics and how these relate to the experiences or cognitive demands of the animal, and the coordination of replay events between different parts of the brain.

%% file: appendices.tex
\section{Construction of distance metric $d$}
\label{app:distance}
We consider the discretised environment as a graph with discrete positions constituting the nodes and an edge connects every pair of nodes for which the corresponding positions are adjacent horizontally, vertically or diagonally. Edges are weighted by the distance between the centroids of the corresponding positions. Then we define $d \lb x', x'' \rb$ as the sum of the weights of the edges that form the shortest path from $x'$ to $x''$. The shortest paths between every pair of nodes on the graph can be computed efficiently using the Floyd-Warshall algorithm or Johnson's algorithm (\cite{Leiserson2001}).

\section{Posterior parameter sampling distributions}
\subsection{Spike train model parameters}
\label{subapp:posteriorlambda}
We consider the posterior distribution at time step $t$ for parameter $\lambda_{i, n}$. We have
\begin{align}
&p \lb   \lambda_{i, n}   \mid   x_{1:t},   \mathbf{y}_{1:t},   s_{0:t},   \mathbf{\theta},   \kappa,   \mathbf{\phi}   \rb   \nonumber\\
&\qquad \propto   p \lb   \mathbf{y}_{1:t}   \mid   s_{0:t},   \mathbf{\theta}   \rb
              p \lb   \lambda_{i, n}   \mid   \mathbf{\phi},   \kappa   \rb   \nonumber\\
&\qquad \propto   \lb   \prod_{u \le t: s_u = i}   \frac{   \exp \lcb   -\delta t   \lambda_{i, n}   \rcb   }{   y_{u, n}!   }
              \lb   \delta t   \lambda_{i, n}   \rb^{y_{u, n}}   \rb   \nonumber
\lambda_{i, n}^{\alpha - 1}   \exp \lcb   -\lambda_{i, n}   \beta   \rcb    \nonumber\\
&\qquad \propto   \lb   \prod_{u \le t: s_u = i}   \exp \lcb   -\delta t   \lambda_{i, n}   \rcb
              \lambda_{i, n}^{y_{u, n}}   \rb   \lambda_{i, n}^{\alpha - 1}   \nonumber
\exp \lcb   -\lambda_{i, n}  \beta   \rcb   \nonumber\\
&\qquad =   \exp \lcb   -\lb   \delta t   c_{i, t}   +   \beta   \rb   \lambda_{i, n}   \rcb   \nonumber
\lambda_{i, n}^{\sum_{u \le t: s_u = i}   y_{u, n}   +   \alpha   -   1},
\end{align}
which is, up to a normalising constant, the pdf of $\texttt{Gam}\lb \lambda_{i, n}; \alpha^*, \beta^* \rb$ with shape, rate parameterisation.

\subsection{Position model parameters}
\label{subapp:posteriormean}
We first state some properties of the transformation $\mathbf{f}_x$. We have
\begin{align}
\mathbf{f}_{x'} \lb   x''   \rb   =   &   - \mathbf{f}_{x''} \lb   x'   \rb   \qquad   \text{and}\\
\mathbf{f}_{x'} \lb   x''   \rb   =   &   \mathbf{f}_{x'} \lb   x'''   \rb   +   \mathbf{f}_{x'''} \lb   x''   \rb.
\end{align}
Both are properties of vectors in $\mathbb{R}^2$. From Eqs. \eqref{eq:posteriorofmean1} and \eqref{eq:posteriorofmean2} we have
\begin{align}
&   p \lb   \xi_i   \mid x_{1:t},   \mathbf{y}_{1:t},   s_{0:t},   \mathbf{\theta},   \kappa,   \mathbf{\phi}   \rb   \nonumber\\
&   \propto   \exp \lcb   \sum_{u \le t: s_u = i}   \mathbf{f}_{\xi_i} \lb   x_u   \rb^\intercal
\Sigma_i^{-1}   \mathbf{f}_{\xi_i} \lb   x_u   \rb   \rcb   \nonumber\\
&   =   \exp \lcb   \sum_{u \le t: s_u = i}   \lb   \mathbf{f}_{\xi^*} \lb   x_u   \rb
-   \mathbf{f}_{\xi^*} \lb   \xi_i   \rb   \rb^\intercal
\Sigma_i^{-1}   \lb   \mathbf{f}_{\xi^*} \lb   x_u   \rb
-   \mathbf{f}_{\xi^*}   \lb   \xi_i   \rb   \rb   \rcb,
\end{align}
in which $^\intercal$ denotes the transpose operator. The exponent expands as
\begin{equation}
c_{i, t}   \mathbf{f}_{\xi^*} \lb   \xi_i   \rb^\intercal
\Sigma_i^{-1}   \mathbf{f}_{\xi^*} \lb   \xi_i   \rb
-   2c_{i, t}   \mathbf{f}_{\xi^*} \lb   \xi_i   \rb
\Sigma_i^{-1}   c_{i, t}^{-1}
\sum_{u \le t: s_u = i}   \mathbf{f}_{\xi^*} \lb   x_u   \rb
+   \sum_{u \le t: s_u = i}   \mathbf{f}_{\xi^*} \lb   x_u   \rb^\intercal
\Sigma_i^{-1}   \mathbf{f}_{\xi^*} \lb   x_u   \rb.
\end{equation}
Now we obtain the form of the Gaussian posterior by completing the square. The exponent becomes
\begin{align}
&   c_{i, t}   \lb   c_{i, t}^{-1}   \sum_{u \le t: s_u = i}   \mathbf{f}_{\xi^*} \lb   x_u   \rb
-   \mathbf{f}_{\xi^*} \lb   \xi_i   \rb   \rb^\intercal
\Sigma_i^{-1}   \lb   c_{i, t}^{-1}
\sum_{u \le t: s_u = i}   \mathbf{f}_{\xi^*} \lb   x_u   \rb
-   \mathbf{f}_{\xi^*} \lb   \xi_i   \rb   \rb   \nonumber\\
&\qquad   +   \sum_{u \le t: s_u = i}   \mathbf{f}_{\xi^*} \lb   x_u   \rb^\intercal
\Sigma_i^{-1}   \mathbf{f}_{\xi^*} \lb   x_u   \rb   \nonumber\\
&\qquad   -   c_{i, t}   \lb   c_{i, t}^{-1}   \sum_{u \le t: s_u = i}   \mathbf{f}_{\xi^*} \lb   x_u   \rb   \rb^\intercal
\Sigma_i^{-1}   \lb   c_{i, t}^{-1}   \sum_{u \le t: s_u = i}   \mathbf{f}_{\xi^*} \lb   x_u   \rb   \rb,
\end{align}
but the last two terms do not depend on $\xi_i$ and so the posterior is, up to a normalising constant,
\begin{align}
&   \exp \lcb   c_{i, t}   \lb   c_{i, t}^{-1}   \sum_{u \le t: s_u = i}   \mathbf{f}_{\xi^*} \lb   x_u   \rb   -   \mathbf{f}_{\xi^*} \lb   \xi_i   \rb   \rb^\intercal
\Sigma_i^{-1}   \lb   c_{i, t}^{-1}   \sum_{u \le t: s_u = i}   \mathbf{f}_{\xi^*} \lb   x_u   \rb   -   \mathbf{f}_{\xi^*} \lb   \xi_i   \rb   \rb   \rcb   \nonumber\\
&   =   \exp \lcb   c_{i, t}   \lb   \mathbf{f}_{\xi^*} \lb   \xi_i   \rb \rb^\intercal
\Sigma_i^{-1}   \lb   \mathbf{f}_{\xi^*} \lb   \xi_i   \rb   \rb   \rcb,
\end{align}
if we choose $\xi^* = \bar{x}_i$, where $\bar{x}_i$ satisfies $c_{i, t}^{-1}   \sum_{u \le t: s_u = i}   \mathbf{f}_{\bar{x}_i} \lb   x_u   \rb   =   0$. In practise there may not be a solution due to the discretisation of space, so we take a value for $\bar{x}_i$ that minimises this expression as per Eq. \eqref{eq:posteriorgaussian}.

\subsection{Rows of the transition matrix}
\label{subapp:transitionmatrows}
We use the algorithm of \cite{Wong1998} to sample $\mathbf{P}_{i, \cdot}$ from the Generalised Dirichlet distribution with parameter vectors $\mathbf{\zeta}_i, \mathbf{\gamma}_i$. The Generalised Dirichlet distribution can be constructed as a product of Beta distributions with parameters $\zeta_{i, j}, \eta_{i, j}$ for $1 \le j \le \kappa$, from which the $\mathbf{\gamma}_i$ parameters can be derived as:
\begin{align}
\gamma_{i, \kappa}   =   &   \eta_{i, \kappa}   -   1,   \nonumber\\
\gamma_{i, j}   =   &   \eta_{i, j}   -   \zeta_{i, j + 1}   -   \eta_{i, j + 1}   \qquad   \text{for }   j   =   \kappa - 1, \dots, 1
\end{align}
(for details see \cite{Wong1998}). Therefore, if we set
\begin{align}
\eta_{i, \kappa}   =   &   \gamma_{i, \kappa}(t)   +   1,   \nonumber\\
\eta_{i, j}   =   &   \gamma_{i, j}(t)   +    \zeta_{i, j + 1}   +   \eta_{i, j + 1}   \qquad   \text{for }   j   =   \kappa - 1, \dots, 1,
\end{align}
we retrieve the parameters of the underlying Beta distributions, and we can use the following procedure to sample $\mathbf{P}_{i, \cdot}$:
\begin{itemize}
\item sample $P_{i, 1}   \sim   \texttt{Beta} \lb   \zeta_{i, 1},   \eta_{i, 1}   \rb$
\item set $\sigma   \gets   P_{i, 1}$
\item for $j$ from $2$ to $\kappa$:
\begin{itemize}
\item sample $P_{i, j}   \sim   \texttt{Beta} \lb   \zeta_{i, j},   \eta_{i, j}   \rb$
\item then $P_{i, j}   \gets   P_{i, j}   \lb   1   -   \sigma   \rb$
\item set $\sigma   \gets   \sigma   +   P_{i, j}$
\end{itemize}
\end{itemize}

\section{Viterbi-like algorithm for decoding position}
\label{app:viterbiforposition}
Here is described a recursive algorithm to find
\begin{equation}
\hat{x}_{1:T}
=   \argmax_{x_{1:T}}   p \lb   x_{1:T},   \mathbf{y}_{1:T}, \theta   \rb
=   \argmax_{x_{1:T}}   p \lb   x_{1:T}   \mid   \mathbf{y}_{1:T}, \theta   \rb.
\end{equation}
First, define
\begin{equation}
V_t \lb   v, j   \rb
\defeq   \max_{x_{1:t - 1}}   \lcb   p \lb   S_t = j,   X_{1:t - 1} = x_{1:t - 1},   X_t = v,   \mathbf{y}_{1:t},   \theta   \rb   \rcb.
\end{equation}
Now notice that
\begin{align}
V_t \lb   v, j   \rb
=   &\max_{x_{1:t - 1}}   \lcb
\sum_{i = 1}^\kappa   p \lb   S_t = j   \mid   S_{t - 1} = i,   \theta   \rb \right.\nonumber\\
&\qquad\left.   \times   p \lb   S_{t - 1} = i,   X_{1:t - 2} = x_{1:t - 2},   X_{t - 1} = x_{t - 1},   \mathbf{y}_{1:t - 1}   \mid   \theta   \rb   \vphantom{\sum_{i = 1}^\kappa} \rcb \nonumber\\
&\times   p \lb   \mathbf{y}_t   \mid   S_t = j,   \theta   \rb
p \lb   X_t = v   \mid   S_t = j,   \theta   \rb,
\end{align}
by the conditional independence structure and since the last two terms do not depend on $x_{1:t - 1}$. This suggests the recursions
\begin{align}
V_t \lb   v, j   \rb
=   &\max_{u}    \lcb
\sum_{i = 1}^\kappa   p \lb   S_t = j   \mid   S_{t - 1} = i,   \theta   \rb   V_{t - 1} \lb   u, i   \rb   \rcb \nonumber\\
&\times   p \lb   \mathbf{y}_t   \mid   S_t = j,   \theta   \rb
p \lb   X_t = v   \mid   S_t = j,   \theta   \rb,
\end{align}
for $t$ from $2$ to $T$, with initialisation
\begin{align}
V_1 \lb   v, j   \rb
=   &\sum_{i = 1}^\kappa   p \lb   S_1 = j   \mid  S_0 = i   \rb   p \lb   S_0 = i   \mid   \theta   \rb \nonumber\\
&\times   p \lb   \mathbf{y}_1   \mid   S_1 = j,   \theta   \rb   p \lb   X_1 = v   \mid   S_1 = j,   \theta   \rb.
\end{align}
Once these have been computed for each $v \in \lcb   1, 2, \dots, M   \rcb$ and each $j \in \lcb   1, 2, \dots, \kappa   \rcb$ we can use the recursions
\begin{equation}
\hat{x}_t   =   \argmax_{v}   \sum_{j = 1}^\kappa   p \lb   X_{t + 1} = \hat{x}_{t + 1}   \mid   S_{t + 1} = j,   \theta   \rb
\sum_{i = 1}^\kappa   p \lb   S_{t + 1} = j   \mid   S_t = i,   \theta   \rb   V_t \lb   v, i   \rb
\end{equation}
for $t$ from $T - 1$ to $1$, with initialisation
\begin{equation}
\hat{x}_T   =   \argmax_{v}   \sum_{j = 1}^\kappa   V_T \lb   v, j   \rb.
\end{equation}

\section{Algorithm for computing the posterior probability of a trajectory}
\label{app:trajectoryprobability}
Here is described an efficient recursive algorithm for computing the posterior probability of a position trajectory $x_{1:a}$ of length $a$ at any offset $t$; that is
\begin{equation}
p \lb   X_t = x_1, X_{t + 1} = x_2, \dots, X_{t + a - 1} = x_a   \mid   \mathbf{y}_{1:T},   \theta   \rb
\end{equation}
for any $t \in \lcb   1, 2, \dots, T - a + 1   \rcb$.

We begin by performing the forward and backward recursions; then the algorithm has two stages. First we compute the intermediary quantities
\begin{equation}
p \lb   S_{t + a - 1} = i,   X_t = x_1, X_{t + 1} = x_2, \dots, X_{t + a - 1} = x_a,   \mathbf{y}_{1:t + a - 1}   \mid   \theta   \rb
\end{equation}
for each $i \in \lcb   1, 2, \dots, \kappa   \rcb$ using the forward accumulation steps
\begin{align}
&p \lb   S_{t + u - 1} = j,   X_t = x_1, \dots, X_{t + u - 1} = x_u,   \mathbf{y}_{1:t + u - 1}   \mid   \theta   \rb \nonumber\\
&\qquad =   p \lb   X = x_u   \mid   S = j,   \theta   \rb   p \lb   \mathbf{y}_{t + u - 1}   \mid   S_{t + u - 1} = j,   \theta   \rb \nonumber\\
&\qquad \quad\times   \sum_{i = 1}^\kappa   p \lb   S_{t + u - 1} = j   \mid   S_{t + u - 2} = i,   \theta   \rb \nonumber\\
&\qquad \quad\times   p \lb   S_{t + u - 2} = i,   X_t = x_1, \dots, X_{t + u - 2} = x_{u - 1},   \mathbf{y}_{1:t + u - 2}   \mid   \theta   \rb
\end{align}
for $u$ from $2$ to $a$ with initialisation
\begin{equation}
p \lb   S_t = j,   X_t = x_1,   \mathbf{y}_{1:t}   \mid   \theta   \rb
=   p \lb   X = x_1   \mid   S = j,   \theta   \rb   p \lb   S_t = j,   \mathbf{y}_{1:t}   \mid   \theta   \rb,
\end{equation}
where $p \lb   S_t = j,   \mathbf{y}_{1:t}   \mid   \theta   \rb$ is the $t$th forward function evaluated at state $j$. Then we perform the second stage:
\begin{align}
&p \lb   X_t = x_1, X_{t + 1} = x_2, \dots, X_{t + a - 1} = x_a   \mid   \mathbf{y}_{1:T},   \theta   \rb \nonumber\\
&\propto \sum_{j = 1}^\kappa   p \lb   S_{t + a - 1} = i,   X_t = x_1, X_{t + 1} = x_2, \dots, X_{t + a - 1} = x_a,   \mathbf{y}_{1:t + a - 1}   \mid   \theta   \rb \nonumber\\
&\qquad \times p \lb   \mathbf{y}_{t + a:T}   \mid   S_{t + a - 1} = j,   \theta   \rb,
\end{align}
in which $p \lb   \mathbf{y}_{t + a:T}   \mid   S_{t + a - 1} = j,   \theta   \rb$ is the $\lb T - t - a + 1 \rb$th backward function evaluated at state $j$, and with normalising constant
\begin{equation}
p \lb   \mathbf{y}_{1:T}   \mid   \theta   \rb
=   \sum_{j = 1}^\kappa   p \lb   S_T = j,   \mathbf{y}_{1:T}   \mid   \theta   \rb.
\end{equation}
For computing the above over all possible $t$, the time and memory requirements are proportional to those of the forward-backward algorithm.

\section{Algorithm for computing the marginal probability of a trajectory}
\label{app:marginaltrajectoryprobability}

Here we describe a recursive algorithm for computing the marginal probability of a position trajectory $x_{1:a}$ of length $a$ at any offset $t$,
\begin{equation}
p \lb   X_t = x_1, X_{t + 1} = x_2, \dots, X_{t + a - 1} = x_a   \mid   \theta   \rb
\end{equation}
for any $t \in \lcb   1, 2, \dots, T - a + 1   \rcb$. Using the model conditional distribution over positions given state and the state transition matrix, we recursively compute the quantities
\begin{align}
&p \lb   S_{t + u - 1} = j,   X_t = x_1, X_{t + 1} = x_2, \dots, X_{t + u - 1} = x_u   \mid   \theta   \rb \nonumber\\
&\quad =   \sum_{i = 1} ^ \kappa   
P_{i, j}   p \lb   X_{t + u - 1} = x_u   \mid   S_{t + u - 1} = j,   \theta   \rb \nonumber\\
&\qquad \times   p \lb   S_{t + u - 2} = i,   X_t = x_1, X_{t + 1} = x_2, \dots, X_{t + u - 2} = x_{u - 1}   \mid   \theta   \rb
\end{align}
for each $j \in \lcb   1, 2, \dots, \kappa   \rcb$ and for $u$ from 2 to $a$. We assume (as discussed in Section \ref{subsubsec:positiondecoding}) that by $t$ the Markov chain has reached its equilibrium distribution $\nu$ so that we can initialise the algorithm with
\begin{equation}
p \lb   S_t = j,   X_t = x_1   \mid   \theta   \rb
=   \nu_j   p \lb   X_t = x_1   \mid   S_t = j,   \theta   \rb
\end{equation}
for each $j \in \lcb   1, 2, \dots, \kappa   \rcb$. After performing the above recursions, we obtain the desired quantity with:
\begin{align}
&p \lb   X_t = x_1, X_{t + 1} = x_2, \dots, X_{t + a - 1} = x_a   \mid   \theta   \rb \nonumber\\
&\qquad =   \sum_{i = 1} ^ \kappa   p \lb   S_{t + a - 1} = i,   X_t = x_1, X_{t + 1} = x_2, \dots, X_{t + a - 1} = x_a   \mid   \theta   \rb.
\end{align}